\documentclass[runningheads]{llncs}

\usepackage[T1]{fontenc}
\usepackage{amsfonts}
\usepackage{paralist}

\usepackage{tikz}
\usepackage{amsmath}
\usepackage{tabularray}
\usepackage{multirow}
\usepackage{adjustbox}
\usepackage{color}
\usepackage{graphicx,wrapfig,lipsum}
\usepackage{amssymb}
\usepackage[font=small,skip=3pt]{caption}
\usepackage[shortlabels]{enumitem}
\usepackage[ruled,vlined]{algorithm2e}
\usepackage{algpseudocode}
\usepackage{tcolorbox}
\newcommand{\overbar}[1]{\mkern 1.5mu\overline{\mkern-1.5mu#1\mkern-1.5mu}\mkern 1.5mu}

\usepackage[colorlinks,pagebackref,breaklinks]{hyperref} 

\hypersetup{
  hypertexnames=true, linkcolor=blue, anchorcolor=black,
  citecolor=magenta, urlcolor=alizarin  
}

\usepackage{color, colortbl}
\definecolor{Gray}{gray}{0.9}
\definecolor{airforceblue}{rgb}{0.36, 0.54, 0.66}
\definecolor{aliceblue}{rgb}{0.94, 0.97, 1.0}
\definecolor{alizarin}{rgb}{0.82, 0.1, 0.26}
\definecolor{amber}{rgb}{1.0, 0.75, 0.0}
\definecolor{amber(sae/ece)}{rgb}{1.0, 0.49, 0.0}
\definecolor{arsenic}{rgb}{0.23, 0.27, 0.29}
\definecolor{bronze}{rgb}{0.8, 0.5, 0.2}
\definecolor{battleshipgrey}{rgb}{0.52, 0.52, 0.51}
\definecolor{bole}{rgb}{0.47, 0.27, 0.23}
\definecolor{bulgarianrose}{rgb}{0.28, 0.02, 0.03}
\definecolor{cadet}{rgb}{0.33, 0.41, 0.47}
\definecolor{ceil}{rgb}{0.57, 0.63, 0.81}
\definecolor{cerulean}{rgb}{0.0, 0.48, 0.65}
\definecolor{charcoal}{rgb}{0.21, 0.27, 0.31}
\definecolor{coolblack}{rgb}{0.0, 0.18, 0.39}
\definecolor{coolgrey}{rgb}{0.55, 0.57, 0.67}
\definecolor{darkcandyapplered}{rgb}{0.64, 0.0, 0.0}
\definecolor{darkbrown}{rgb}{0.4, 0.26, 0.13}
\definecolor{darkcerulean}{rgb}{0.03, 0.27, 0.49}
\definecolor{darkgray}{rgb}{0.66, 0.66, 0.66}
\definecolor{darkjunglegreen}{rgb}{0.1, 0.14, 0.13}
\definecolor{darktaupe}{rgb}{0.28, 0.24, 0.2}
\definecolor{davy\'sgrey}{rgb}{0.33, 0.33, 0.33}
\definecolor{frenchblue}{rgb}{0.0, 0.45, 0.73}
\definecolor{almond}{rgb}{0.94, 0.87, 0.8}
\definecolor{beaublue}{rgb}{0.74, 0.83, 0.9}
\definecolor{beige}{rgb}{0.96, 0.96, 0.86}
\definecolor{bisque}{rgb}{1.0, 0.89, 0.77}
\definecolor{black}{rgb}{0.0, 0.0, 0.0}
\definecolor{fluorescentorange}{rgb}{1.0, 0.75, 0.0}
\definecolor{ghostwhite}{rgb}{0.97, 0.97, 1.0}
\definecolor{antiquewhite}{rgb}{0.98, 0.92, 0.84}

\makeatletter
\newcommand{\printfnsymbol}[1]{%
  \textsuperscript{\@fnsymbol{#1}}%
}
\makeatother

\begin{document}
%
\title{Split Without a Leak: Reducing Privacy Leakage in Split Learning\thanks{\protect\footnotesize This work was funded by the Technology Innovation Institute (TII) for the project ARROWSMITH and from Horizon Europe for HARPOCRATES (101069535).}}

\titlerunning{Split Without a Leak}
%
\author{Khoa Nguyen\inst{1}\orcidID{0000-0001-5089-5250} \and
Tanveer Khan\inst{1}\orcidID{0000-0001-7296-2178}\thanks{The first two authors contributed equally to this work.} \and
Antonis Michalas\inst{1,2}\orcidID{0000-0002-0189-3520}}

\authorrunning{K. Nguyen et al.}

\institute {Tampere University, Finland \and RISE Research Institutes of Sweden \\
\email{\{khoa.nguyen,tanveer.khan, antonios.michalas\}@tuni.fi}}

\maketitle              
\begin{abstract}
The popularity of Deep Learning (DL) makes the privacy of sensitive data more imperative than ever. As a result, various privacy-preserving techniques have been implemented to preserve user data privacy in DL. Among various privacy-preserving techniques, collaborative learning techniques, such as Split Learning (SL) have been utilized to accelerate the learning and prediction process. Initially, SL was considered a promising approach to data privacy. However, subsequent research has demonstrated that SL is susceptible to many types of attacks and, therefore, it cannot serve as a privacy-preserving technique. Meanwhile, countermeasures using a combination of SL and encryption have also been introduced to achieve privacy-preserving deep learning. In this work, we propose a hybrid approach using SL and Homomorphic Encryption (HE). The idea behind it is that the client encrypts the activation map (the output of the split layer between the client and the server) before sending it to the server. Hence, during both forward and backward propagation, the server cannot reconstruct the client’s input data from the intermediate activation map. This improvement is important as it reduces privacy leakage compared to other SL-based works, where the server can gain valuable information about the client’s input. In addition, on the MIT-BIH dataset, our proposed hybrid approach using SL and HE yields faster training time (about~6 times) and significantly reduced communication overhead (almost~160 times) compared to other HE-based approaches, thereby offering improved privacy protection for sensitive data in DL.


\end{abstract}
\section{Introduction}
\label{sec:intro}




Machine learning (ML) 
can boost productivity in various fields by leveraging data to make predictions and decisions without explicit human instructions. However, the data should be of excellent quality in order to achieve good results after training 
ML models. 
Any ML algorithm only performs well when it is provided with a large amount of high-quality training data~\cite{chai2022data}. In a nutshell, one can say that data drives ML. Organizations that create and gather data can build and train their ML models. This allows them to make use of their models and provide 
ML as a service 
(MLaaS)~\cite{ribeiro2015mlaas} to 
other organizations. This is beneficial for 
organizations that are unable to build their own models but wish to use this 
method for data predictions. 
However, a cloud-hosted model 
poses 
numerous privacy issues~\cite{hesamifard2018privacy,riazi2019deep,topol2019high}. In order to use it, 
external organizations 
must either upload their input data or download the model. Regarding privacy, uploading input data can be problematic, 
while downloading the model might not be a suitable option due to intellectual property.

A possible solution is to use 
privacy-preserving techniques~\cite{boulemtafes2020review} -- to encrypt the model and the data in a way that 
allows one organization to use a model held by another organization without either party disclosing their intellectual property. 
Homomorphic Encryption (HE)~\cite{acar2018survey} and Secure Multi-Party Computation (SMPC)~\cite{lindell2020secure} are the 
best-known 
techniques 
enabling computations over encrypted data, hence preserve input privacy. Both of these techniques seem like a silver bullet solution because they enable computation on encrypted data without revealing the underlying information. However, some issues lie behind the surface that can make their implementation more difficult, such as communication cost in SMPC and computation complexity in HE~\cite{cabrero2021sok}.

In addition to HE and SMPC, collaborative learning techniques like Federated Learning (FL)~\cite{bonawitz2019towards} and Split Learning (SL)~\cite{gupta2018distributed} can also be used to preserve user data privacy. Both FL and SL are similar in that they both allow for the training of ML models on large distributed datasets, \textit{without} the need to centralize the data. However, they differ in how the models are split and trained, and in the types of applications they are used for.  
FL is 
used to train a ML model on a large number of devices, each 
using local data. In FL, each device trains a local copy of the model, and the models are then combined on a central server. Through 
this approach 
the model is trained on a large amount of non-centralized data. This 
can be useful in the case of sensitive data 
or devices 
not connected to the internet. SL is a distributed deep learning technique that splits a neural network into two parts, first half 
directed to the client(s) and second to the server. Both the client(s) and the server collaboratively train the split model, though they \textit{cannot} access each other's parts, 
SL provides multiple benefits such as: 
\textit{(i)} joint neural network training by multiple parties, such as a client and a server, 
in which 
each party is protected from disclosing their parts of the model, \textit{(ii)} 
ML model training by users. In this case, users do not share 
their raw data with a server running part of a Deep Neural Network Model (DNN), hence preserving 
their privacy, \textit{(iii)} 
computational burden reduction for the client, as SL 
does not run 
the entire model (i.e.\ 
it utilizes a smaller number of layers), and \textit{(iv)} model classification accuracy that is comparable to 
non-split model. Also SL offers several significant advantages over FL, such as~\cite{singh2019detailed,poirot2019split,turina2021federated}
\begin{inparaenum}[\it (i)] 
\item \textit{not} full disclosure of the model's architecture and weights \item 
peak performance 
through 
minimal resource consumption \item 
required bandwidth reduction \item 
reduced computational burden on the data owners’ devices. \end{inparaenum}

By definition, SL provides an additional layer of privacy protection, however, Abudhabba \textit{et al.}~\cite{abuadbba2020can} showed that adding SL to one-dimensional (1D) Convolutional Neural Network (CNN) models for time-series data like ECG signals could result in significant privacy leakage. To reduce privacy leakage, the authors used two more mitigation techniques, namely 
differential privacy~\cite{Michalas:22:CODASPY:PrivateLives} and 
additional layers. However, based on the findings, none of these techniques can significantly reduce privacy leakage from 
the channels of the SL activation. Additionally, the accuracy of the joint model is greatly decreased by both 
methods.

While there is a plethora of Privacy-preserving ML (PPML) works utilizing HE to protect users' input, 
to the best of our knowledge, until now there is only one work~\cite{khan2023split} that combines HE with SL. In this work, we concentrate on collaborative client-server training of an ML model in a privacy-preserving manner by  constructing a model that uses HE to reduce privacy leakage in SL. 


\subsection{Contributions}
	\begin{enumerate}[\bf C1.]
	    \item In~\cite{khan2023split}, the authors proposed an SL protocol with HE. However, 
        by analyzing the proposed protocol, we found that the server can gain some valuable information on the client's input data by exploiting the gradients sent from the client in the backward pass of the protocol. This, results in important privacy leakage. In this paper, we propose an improved protocol to face this privacy leakage.
        \item 
        We took the U-shaped split 1D CNN proposed in~\cite{khan2023split}, tested this model with plaintext activation maps, and an encrypted version using HE to allow the server to perform computations on encrypted activation maps. 
        Even though the 1D CNN model is identical to that of~\cite{khan2023split}, our proposed protocol optimizes the amount of information being transferred between the client and the server, through that not only solve the privacy leakage in~\cite{khan2023split}, but also improve the communication cost of the whole protocol.
	    \item We have designed a new training protocol permitting 
        the extensive use of CKKS HE scheme's packing feature~\cite{cheon2017homomorphic}. Combining packing and optimization techniques in C2, our training becomes much faster (almost~6 times) compared to~\cite{khan2023split} and also requires much less overhead communication (about~160 times) on the MIT-BIH dataset.
	\end{enumerate}
 
\subsection{Organization}
The rest of the paper is organized as follows: 	In~\autoref{sec:relatedworks}, we present important published works in the area of SL. 
In~\autoref{sec:preliminaries}, we provide the necessary background information about 1D CNN, HE and SL. 
The architectures of the local and split models on plaintext data are presented in~\autoref{sec:network}. In~\autoref{sec:privacyLeakageAnal}, we report the privacy leakage we noted in the paper~\cite{khan2023split} while  \autoref{sec:splitmodel} presents the design and implementation of the split 1D CNN training protocol on encrypted data that solves this privacy leakage. Then,~\autoref{sec:performance}
refers extensively to experimental results and finally the paper is concluded in~\autoref{sec:conclusion}.

\section{Related Works}
\label{sec:relatedworks}

SL has been proposed as a privacy-preserving implementation of collaborative learning~\cite{vepakomma2018split,vepakomma2019reducing,poirot2019split} and has gained particular interest due to its efficiency and simplicity. For user data privacy, SL relies on the fact that only intermediate activation maps are shared 
between the parties. It 
can 
enable more privacy-preservation 
and efficient use of deep learning models in a variety of settings~\cite{gupta2018distributed}. This is why 
SL-based solutions have been implemented and adopted in commercial as well as open-source applications. For example, SL 
could enable medical image 
analysis on a patient's device, while 
securing their privacy and sensitive data~\cite{vepakomma2018split}. In addition, SL can 
efficiently process data generated by IoT devices, by keeping the computationally intensive part of a model on a remote server~\cite{koda2019one,yansong2020end,lim2020incentive,samikwa2022ares}.

Although the previous studies~\cite{gupta2018distributed,vepakomma2018split,poirot2019split} assumed that SL is designed to protect the intellectual property of the shared model and to reduce the risk of inference attacks\footnote{A type of attack where an attacker can infer only specific properties of the private training instances, rather than reconstructing the entire input.} perpetrated by a malicious server, recent studies have shown that these assumptions are false and 
privacy leakage risks do exist in SL. In~\cite{vepakomma2019reducing}, the authors analyzed the privacy leakage of SL and found a considerable leakage from the split layer in the 2D~CNN model. Also, the authors in paper~\cite{abuadbba2020can} performed experiments on the 1D~CNN model and found that sharing the activation from the split layer results in severe privacy leakage. Also, the paper described various structural vulnerabilities of SL and showed how to exploit them and violate
the protocol’s privacy-preserving property.
This makes SL a prime target for various types of attacks such as model inversion attack~\cite{he2019model,abuadbba2020can,titcombe2021practical} and property inference attack~\cite{pasquini2021unleashing}.

Various SL privacy threats and attacks have been presented. 
In addition, countermeasures have also
been introduced to achieve privacy-preserving deep learning.
For example, Vepakomma \textit{}et al.~\cite{vepakomma2019reducing} proposed a method for limiting data reconstruction in split neural networks (SplitNN) by minimizing the distance correlation between the input data and the intermediate tensors during model training. Also, to protect the SL from property inference attack, one approach is to use secure aggregation protocol that can protect the privacy of the clients' data while still allowing for collaborative learning~\cite{pasquini2021unleashing}. Another approach is to use differential privacy techniques to add noise, which can help prevent attackers from inferring sensitive information from the model updates~\cite{chen2020differential}. Abuadbba \textit{et al.}~\cite{abuadbba2020can} applies noise to the intermediate tensors in a SplitNN
as a defence mechanism 
against model inversion attacks on one-dimensional ECG data.  A similar method, Shredder~\cite{mireshghallah2020shredder}, was introduced by Mireshghallah \textit{et al.}. 
It adaptively generates a noise mask to minimize mutual information between input and intermediate
data. Also, the authors in~\cite{titcombe2021practical}, proposed a simple additive noise method to defend against model inversion. This
revealed 
that the method can significantly reduce attack efficacy. In addition to the above-mentioned techniques, hybrid approaches 
have also been combined with FL to yield a more scalable privacy-preserving training protocol, such as~\cite{thapa2022splitfed}. 
Pereteanu \textit{et al.,} propose splitting the server network into private sections separated by a public section that can be assessed in plaintext by client to speed up classification while using HE~\cite{pereteanu2022split}. While their approach are somewhat similar to our work, it is limited to client input classification and does not permit a client to customize a network to their private dataset.
Recently, the authors of~\cite{khan2023split} proposed an approach that combines SL and HE that can be considered as the closest work to ours. 
In their model, 
prior to sending the intermediate input to the server, the client first encrypts it and sends the encrypted activation map to the server. However, one of the shortcomings of this hybrid approach, as we show in~\autoref{sec:privacyLeakageAnal}, is that during the backward propagation, by exploiting the gradients sent from the client, the server can gain 
valuable information on the client's input data, 
thus 
causing privacy leakage. 
We propose an improved protocol 
to address 
this privacy leakage~\cite{khan2023split}. Our protocol is faster and incurs less communication overhead.

\section{Preliminaries}
\label{sec:preliminaries}
\subsection{Convolutional Neural Networks}
CNNs are a class of neural network that uses the convolution kernels to slide along the input signal and produce output feature maps. CNNs are used to process grid-like topology data, for example, they were used to recognize handwritten zip codes in 2D images~\cite{lecun89zipCode}. CNNs can also process 1D~\cite{kiranyaz20211d} or 3D signals~\cite{Ji3DCNN}. In this work, we employ a 1D CNN on ECG data to classify heartbeats.

If we have a kernel $\mathbf{w}$ of size $m$, and at every step we stride the kernel $n$ steps, then the 1D convolution layer performs the operation:
$\mathbf{z}_i = \sum_{j=0}^{m-1}\mathbf{w}_j \cdot \mathbf{x}_{i\times n+j}$
Apart from the Conv1D layers, our 1D CNN also consists of other layers: 
\begin{itemize}
    \item \underline{Leaky ReLU (LReLU)}: Leaky Rectify Linear Unit is a non-linear function that can be described as: 
 $   f(x) = 
        \begin{cases}
            x, & \text{if} ~~ x \geq 0 \\
            \alpha x, & \text{otherwise}
        \end{cases}$
    
    where $\alpha$ is called the "negative slope" -- a small number such as 0.01.
    \item \underline{Max Pooling}: An operation that calculates maximum value in each patch of a feature map. The result is a down-sampled feature map that contains the maximum values that highlight important features of those patches.
    \item \underline{Fully Connected (FC) or Linear Layer}: An FC layer performs a weighted affine transformation to its input:
        $\mathbf{y} = \mathbf{x} W^T + \mathbf{b}.$
    \item \underline{Softmax}: The Softmax function is used to transform a vector~$\mathbf{z}~=~(z_1, z_2, \ldots, z_m)$ into a vector of probabilities:
 $\text{Softmax}(z_i) = \frac{e^{z_i}}{\sum_{j=1}^{m} e^{z_j}}$
\end{itemize}
By stacking these layers on top of each other, we construct our 1D CNN as a feature detector and classifier for the ECG datasets. The 1D CNN model can be written as a function $f_{\Theta}$, where $\Theta$ is a set of adjustable (or learnable) parameters. $\Theta$ is first initialized to small random values in the range $[-1, 1]$. 
The training process consists of two phases named forward propagation and backward propagation. More specifically, during forward propagation, an input heartbeat $x$ is propagated forward through the Conv1D layer, pooling layer and fully connected layer to obtain the final output value $\hat{y}$. As for each $x$, there is a corresponding encoded label vector $y$ that represents its ground truth class. 
We aim to find the best set of parameters $\Theta$ that maps $x$ to a predicted output vector $\hat{y} $, where $\hat{y}$ is as closed as possible to $y$ which is measured by a loss (or distance) function $\mathcal{L}(\hat{y}, y)$. $\mathcal{L}$ is chosen to be cross entropy loss in our work. To reach the minimized loss function, we update $\Theta$ using backward propagation. Backward propagation moves from the network's output layer back to the input layer to calculate the gradients of the loss function $\mathcal{L}$ w.r.t the weights $\Theta$. Then, these weights are updated according to the gradients. The full process of calculating the predicted output, the loss function and the gradients, followed by updating the weights on one batch of data is called an ``iteration''. In this paper, we train the neural network with thousands of samples of $x$ and associated $y$. We train the neural network using mini-batch gradient descent with the batch size specified by $n$. The total number of training batches is $N=\frac{|D|}{n}$, where $|D|$ is the size of the dataset. $N$ is equal to the number of training iterations. Once the neural network goes through all the training batches, it has completed one training epoch. This process is repeated for a total of $E$ epochs.  

\subsection{Split Learning}
\begin{figure}[!ht]
\centering
\begin{adjustbox}{width=0.6\textwidth, totalheight=0.35\textwidth}
\tikzset{every picture/.style={line width=0.75pt}}        

\begin{tikzpicture}[x=0.75pt,y=0.75pt,yscale=-1,xscale=1]

\draw  [dash pattern={on 0.84pt off 2.51pt}] (15,17) -- (177.5,17) -- (177.5,129) -- (15,129) -- cycle ;
\draw   (64.5,33.6) .. controls (64.5,31.06) and (66.56,29) .. (69.1,29) -- (161.4,29) .. controls (163.94,29) and (166,31.06) .. (166,33.6) -- (166,47.4) .. controls (166,49.94) and (163.94,52) .. (161.4,52) -- (69.1,52) .. controls (66.56,52) and (64.5,49.94) .. (64.5,47.4) -- cycle ;
\draw   (64.5,63.6) .. controls (64.5,61.06) and (66.56,59) .. (69.1,59) -- (161.4,59) .. controls (163.94,59) and (166,61.06) .. (166,63.6) -- (166,77.4) .. controls (166,79.94) and (163.94,82) .. (161.4,82) -- (69.1,82) .. controls (66.56,82) and (64.5,79.94) .. (64.5,77.4) -- cycle ;
\draw   (64.5,93.6) .. controls (64.5,91.06) and (66.56,89) .. (69.1,89) -- (161.4,89) .. controls (163.94,89) and (166,91.06) .. (166,93.6) -- (166,107.4) .. controls (166,109.94) and (163.94,112) .. (161.4,112) -- (69.1,112) .. controls (66.56,112) and (64.5,109.94) .. (64.5,107.4) -- cycle ;
\draw  [dash pattern={on 0.84pt off 2.51pt}] (14,137) -- (176.5,137) -- (176.5,290) -- (14,290) -- cycle ;
\draw   (63.5,153.6) .. controls (63.5,151.06) and (65.56,149) .. (68.1,149) -- (160.4,149) .. controls (162.94,149) and (165,151.06) .. (165,153.6) -- (165,167.4) .. controls (165,169.94) and (162.94,172) .. (160.4,172) -- (68.1,172) .. controls (65.56,172) and (63.5,169.94) .. (63.5,167.4) -- cycle ;
\draw   (63.5,183.6) .. controls (63.5,181.06) and (65.56,179) .. (68.1,179) -- (160.4,179) .. controls (162.94,179) and (165,181.06) .. (165,183.6) -- (165,197.4) .. controls (165,199.94) and (162.94,202) .. (160.4,202) -- (68.1,202) .. controls (65.56,202) and (63.5,199.94) .. (63.5,197.4) -- cycle ;
\draw   (70.5,213.6) .. controls (70.5,211.06) and (72.56,209) .. (75.1,209) -- (152.9,209) .. controls (155.44,209) and (157.5,211.06) .. (157.5,213.6) -- (157.5,227.4) .. controls (157.5,229.94) and (155.44,232) .. (152.9,232) -- (75.1,232) .. controls (72.56,232) and (70.5,229.94) .. (70.5,227.4) -- cycle ;
\draw   (85.5,245.6) .. controls (85.5,243.06) and (87.56,241) .. (90.1,241) -- (138.9,241) .. controls (141.44,241) and (143.5,243.06) .. (143.5,245.6) -- (143.5,259.4) .. controls (143.5,261.94) and (141.44,264) .. (138.9,264) -- (90.1,264) .. controls (87.56,264) and (85.5,261.94) .. (85.5,259.4) -- cycle ;
\draw    (111.5,113) -- (111.5,145) ;
\draw [shift={(111.5,148)}, rotate = 270] [fill={rgb, 255:red, 0; green, 0; blue, 0 }  ][line width=0.08]  [draw opacity=0] (8.93,-4.29) -- (0,0) -- (8.93,4.29) -- cycle    ;
\draw  [dash pattern={on 0.84pt off 2.51pt}] (184,17) -- (378.5,17) -- (378.5,129) -- (184,129) -- cycle ;
\draw   (254.5,33.6) .. controls (254.5,31.06) and (256.56,29) .. (259.1,29) -- (351.4,29) .. controls (353.94,29) and (356,31.06) .. (356,33.6) -- (356,47.4) .. controls (356,49.94) and (353.94,52) .. (351.4,52) -- (259.1,52) .. controls (256.56,52) and (254.5,49.94) .. (254.5,47.4) -- cycle ;
\draw   (255.5,64.1) .. controls (255.5,61.56) and (257.56,59.5) .. (260.1,59.5) -- (352.4,59.5) .. controls (354.94,59.5) and (357,61.56) .. (357,64.1) -- (357,77.9) .. controls (357,80.44) and (354.94,82.5) .. (352.4,82.5) -- (260.1,82.5) .. controls (257.56,82.5) and (255.5,80.44) .. (255.5,77.9) -- cycle ;
\draw   (256.3,93.6) .. controls (256.3,91.06) and (258.36,89) .. (260.9,89) -- (353.2,89) .. controls (355.74,89) and (357.8,91.06) .. (357.8,93.6) -- (357.8,107.4) .. controls (357.8,109.94) and (355.74,112) .. (353.2,112) -- (260.9,112) .. controls (258.36,112) and (256.3,109.94) .. (256.3,107.4) -- cycle ;
\draw  [dash pattern={on 0.84pt off 2.51pt}] (184,137) -- (379.5,137) -- (379.5,290) -- (184,290) -- cycle ;
\draw   (257.5,153.6) .. controls (257.5,151.06) and (259.56,149) .. (262.1,149) -- (354.4,149) .. controls (356.94,149) and (359,151.06) .. (359,153.6) -- (359,167.4) .. controls (359,169.94) and (356.94,172) .. (354.4,172) -- (262.1,172) .. controls (259.56,172) and (257.5,169.94) .. (257.5,167.4) -- cycle ;
\draw   (257.5,183.6) .. controls (257.5,181.06) and (259.56,179) .. (262.1,179) -- (354.4,179) .. controls (356.94,179) and (359,181.06) .. (359,183.6) -- (359,197.4) .. controls (359,199.94) and (356.94,202) .. (354.4,202) -- (262.1,202) .. controls (259.56,202) and (257.5,199.94) .. (257.5,197.4) -- cycle ;
\draw   (264.5,213.6) .. controls (264.5,211.06) and (266.56,209) .. (269.1,209) -- (346.9,209) .. controls (349.44,209) and (351.5,211.06) .. (351.5,213.6) -- (351.5,227.4) .. controls (351.5,229.94) and (349.44,232) .. (346.9,232) -- (269.1,232) .. controls (266.56,232) and (264.5,229.94) .. (264.5,227.4) -- cycle ;
\draw   (188.5,64.6) .. controls (188.5,62.06) and (190.56,60) .. (193.1,60) -- (241.9,60) .. controls (244.44,60) and (246.5,62.06) .. (246.5,64.6) -- (246.5,78.4) .. controls (246.5,80.94) and (244.44,83) .. (241.9,83) -- (193.1,83) .. controls (190.56,83) and (188.5,80.94) .. (188.5,78.4) -- cycle ;
\draw    (311.5,113) -- (311.5,145) ;
\draw [shift={(311.5,148)}, rotate = 270] [fill={rgb, 255:red, 0; green, 0; blue, 0 }  ][line width=0.08]  [draw opacity=0] (8.93,-4.29) -- (0,0) -- (8.93,4.29) -- cycle    ;
\draw  [dash pattern={on 0.84pt off 2.51pt}] (388.5,16) -- (575.5,16) -- (575.5,128) -- (388.5,128) -- cycle ;
\draw   (487.5,46.6) .. controls (487.5,44.06) and (489.56,42) .. (492.1,42) -- (563.4,42) .. controls (565.94,42) and (568,44.06) .. (568,46.6) -- (568,60.4) .. controls (568,62.94) and (565.94,65) .. (563.4,65) -- (492.1,65) .. controls (489.56,65) and (487.5,62.94) .. (487.5,60.4) -- cycle ;
\draw   (488.5,74.6) .. controls (488.5,72.06) and (490.56,70) .. (493.1,70) -- (563.4,70) .. controls (565.94,70) and (568,72.06) .. (568,74.6) -- (568,88.4) .. controls (568,90.94) and (565.94,93) .. (563.4,93) -- (493.1,93) .. controls (490.56,93) and (488.5,90.94) .. (488.5,88.4) -- cycle ;
\draw   (489.5,103.6) .. controls (489.5,101.06) and (491.56,99) .. (494.1,99) -- (563.9,99) .. controls (566.44,99) and (568.5,101.06) .. (568.5,103.6) -- (568.5,117.4) .. controls (568.5,119.94) and (566.44,122) .. (563.9,122) -- (494.1,122) .. controls (491.56,122) and (489.5,119.94) .. (489.5,117.4) -- cycle ;
\draw  [dash pattern={on 0.84pt off 2.51pt}] (388.5,136) -- (575.5,136) -- (575.5,289) -- (388.5,289) -- cycle ;
\draw   (436.5,152.6) .. controls (436.5,150.06) and (438.56,148) .. (441.1,148) -- (533.4,148) .. controls (535.94,148) and (538,150.06) .. (538,152.6) -- (538,166.4) .. controls (538,168.94) and (535.94,171) .. (533.4,171) -- (441.1,171) .. controls (438.56,171) and (436.5,168.94) .. (436.5,166.4) -- cycle ;
\draw   (437.5,182.6) .. controls (437.5,180.06) and (439.56,178) .. (442.1,178) -- (534.4,178) .. controls (536.94,178) and (539,180.06) .. (539,182.6) -- (539,196.4) .. controls (539,198.94) and (536.94,201) .. (534.4,201) -- (442.1,201) .. controls (439.56,201) and (437.5,198.94) .. (437.5,196.4) -- cycle ;
\draw   (444.5,212.6) .. controls (444.5,210.06) and (446.56,208) .. (449.1,208) -- (526.9,208) .. controls (529.44,208) and (531.5,210.06) .. (531.5,212.6) -- (531.5,226.4) .. controls (531.5,228.94) and (529.44,231) .. (526.9,231) -- (449.1,231) .. controls (446.56,231) and (444.5,228.94) .. (444.5,226.4) -- cycle ;
\draw   (459.5,244.6) .. controls (459.5,242.06) and (461.56,240) .. (464.1,240) -- (512.9,240) .. controls (515.44,240) and (517.5,242.06) .. (517.5,244.6) -- (517.5,258.4) .. controls (517.5,260.94) and (515.44,263) .. (512.9,263) -- (464.1,263) .. controls (461.56,263) and (459.5,260.94) .. (459.5,258.4) -- cycle ;
\draw   (189.5,94.6) .. controls (189.5,92.06) and (191.56,90) .. (194.1,90) -- (242.9,90) .. controls (245.44,90) and (247.5,92.06) .. (247.5,94.6) -- (247.5,108.4) .. controls (247.5,110.94) and (245.44,113) .. (242.9,113) -- (194.1,113) .. controls (191.56,113) and (189.5,110.94) .. (189.5,108.4) -- cycle ;
\draw   (190.5,156.6) .. controls (190.5,154.06) and (192.56,152) .. (195.1,152) -- (243.9,152) .. controls (246.44,152) and (248.5,154.06) .. (248.5,156.6) -- (248.5,170.4) .. controls (248.5,172.94) and (246.44,175) .. (243.9,175) -- (195.1,175) .. controls (192.56,175) and (190.5,172.94) .. (190.5,170.4) -- cycle ;
\draw    (210.5,175) .. controls (208.5,261) and (290.5,289) .. (311.5,233) ;
\draw    (210.5,151) -- (210.5,117) ;
\draw [shift={(210.5,114)}, rotate = 90] [fill={rgb, 255:red, 0; green, 0; blue, 0 }  ][line width=0.08]  [draw opacity=0] (8.93,-4.29) -- (0,0) -- (8.93,4.29) -- cycle    ;
\draw   (395.5,46.6) .. controls (395.5,44.06) and (397.56,42) .. (400.1,42) -- (471.4,42) .. controls (473.94,42) and (476,44.06) .. (476,46.6) -- (476,60.4) .. controls (476,62.94) and (473.94,65) .. (471.4,65) -- (400.1,65) .. controls (397.56,65) and (395.5,62.94) .. (395.5,60.4) -- cycle ;
\draw   (396.5,74.6) .. controls (396.5,72.06) and (398.56,70) .. (401.1,70) -- (471.4,70) .. controls (473.94,70) and (476,72.06) .. (476,74.6) -- (476,88.4) .. controls (476,90.94) and (473.94,93) .. (471.4,93) -- (401.1,93) .. controls (398.56,93) and (396.5,90.94) .. (396.5,88.4) -- cycle ;
\draw   (397.5,103.6) .. controls (397.5,101.06) and (399.56,99) .. (402.1,99) -- (471.9,99) .. controls (474.44,99) and (476.5,101.06) .. (476.5,103.6) -- (476.5,117.4) .. controls (476.5,119.94) and (474.44,122) .. (471.9,122) -- (402.1,122) .. controls (399.56,122) and (397.5,119.94) .. (397.5,117.4) -- cycle ;
\draw    (453.5,122) -- (488.96,144.4) ;
\draw [shift={(491.5,146)}, rotate = 212.28] [fill={rgb, 255:red, 0; green, 0; blue, 0 }  ][line width=0.08]  [draw opacity=0] (8.93,-4.29) -- (0,0) -- (8.93,4.29) -- cycle    ;
 
\draw    (520.5,123) -- (493.85,144.14) ;
\draw [shift={(491.5,146)}, rotate = 321.58] [fill={rgb, 255:red, 0; green, 0; blue, 0 }  ][line width=0.08]  [draw opacity=0] (8.93,-4.29) -- (0,0) -- (8.93,4.29) -- cycle    ;

\draw (78,34) node [anchor=north west][inner sep=0.75pt]   [align=left] {\textbf{Input Data}};

\draw (17,58) node [anchor=north west][inner sep=0.75pt]   [align=left] {\textbf{Client}};

\draw (77,148) node [anchor=north west][inner sep=0.75pt]   [align=left] {};

\draw (16,209) node [anchor=north west][inner sep=0.75pt]   [align=left] {\textbf{Server}};

\draw (91.1,247) node [anchor=north west][inner sep=0.75pt]   [align=left] {\textbf{Label}};

\draw (267,35) node [anchor=north west][inner sep=0.75pt]   [align=left] {\textbf{Input Data}};

\draw (192,23) node [anchor=north west][inner sep=0.75pt]   [align=left] {\textbf{Client}};

\draw (189,252) node [anchor=north west][inner sep=0.75pt]   [align=left] {\textbf{Server}};

\draw (193.1,65) node [anchor=north west][inner sep=0.75pt]   [align=left] {\textbf{Label}};

\draw (492.1,46) node [anchor=north west][inner sep=0.75pt]   [align=left] {\textbf{Input Data}};

\draw (408,17) node [anchor=north west][inner sep=0.75pt]   [align=left] {\textbf{Client 1}};

\draw (465.1,247) node [anchor=north west][inner sep=0.75pt]   [align=left] {\textbf{Label}};

\draw (399.1,48) node [anchor=north west][inner sep=0.75pt]   [align=left] {\textbf{Input Data}};

\draw (497,17) node [anchor=north west][inner sep=0.75pt]   [align=left] {\textbf{Client 2}};

\draw (28,294) node [anchor=north west][inner sep=0.75pt]   [align=left] {{\small (a) Simple Vanilla Split }\\{\small  \ \ \ \ \ \ \ \ \ Learning}};

\draw (190,294) node [anchor=north west][inner sep=0.75pt]   [align=left] {{\small (b) Split Learning without Label}\\{\small  \ \ \ \ \ \ \ \ Sharing (U-shaped)}};

\draw (405,292) node [anchor=north west][inner sep=0.75pt]   [align=left] {{\small (c) Split Learning for }\\{\small Vertically Partitioned Data}};

\end{tikzpicture}
\end{adjustbox}
	\caption{Different Split Learning Configurations}
	\label{fig:sl_configurations}
\end{figure}
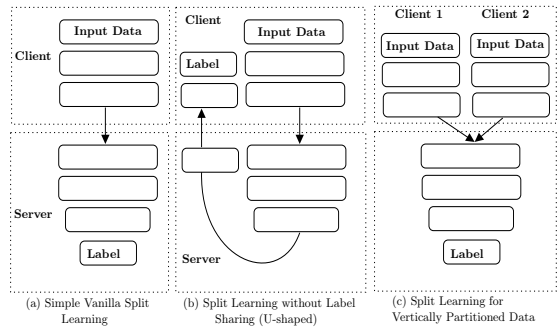

A distributed deep learning method that allows collaboration between different parties to train a model without sharing raw data~\cite{vepakomma2018split}. There are three main configurations of SL, namely vanilla SL, U-shaped SL, and SL for vertically partitioned data~\cite{vepakomma2018split}. These SL configurations are visually demonstrated in~\autoref{fig:sl_configurations}.

We employ the U-shaped SL configuration, where we train the split 1D CNN between the client and the server with the client sharing neither his input data nor the labels. We can see from~\autoref{fig:sl_configurations}(b) that in the forward pass, the client first trains his part of the model until the split layer, then sends the activation maps to the server that 
continues training his part of the model. After finishing 
its part, the server sends its outputs to the client. The client computes the outputs of the network and compares them with the labels to calculate the difference between the predicted outputs and ground-truth labels. In the backward pass, the information flow 
moves in the opposite direction to the forward pass. 

\subsection{Homomorphic Encryption}
With the help of HE, computations on encrypted data can be performed without prior decryption. Homophormism is based on the idea that
when elements are transformed from 
one set to another
, the relationships between them 
are preserved. In 
encryption terms, this means whether operating on encrypted data 
or plaintext data will yield equivalent results. Let's take a homomorphic addition operation as an example. Given two plaintext messages $x_1$, $x_2$ and their corresponding encrypted messages $c_1$, $c_2$. An operation $\oplus$ is an HE addition, if $c_3 = c_1 \oplus c_2$ is the encryption of $x_3 = x_1 + x_2$.

The HE journey started in 1978 
with the paper~\cite{rivest1978data}, in which
the authors envisioned homomorphism as a construction for secure computation and a resort for user data privacy. From then on, we have witnessed the creation of several homomorphic cryptosystems.
Cryptosystems that only enable homomorphic addition or multiplication on ciphertexts (but not both) are called \textit{partial HE}. Examples of partial HE systems are the ElGamal (1985)~\cite{elgamal1985public} or Paillier cryptosystem (1999)~\cite{paillier1999public}. \textit{Fully HE} (FHE) cryptosystems that allow both addition and multiplication on encrypted data, were thought of as impractical until Gentry's first FHE scheme in 2009~\cite{gentry2009fully}. Gentry introduced the concept of bootstrapping, an operation performed to reset the noise level in ciphertexts after 
homomorphic operations. 
Following Gentry's work, 
new HE schemes 
were developed. There are 
two main approaches to new HE schemes, 
namely fast bootstrapping and leveled. Schemes that belong to the
\textit{fast bootstrapping FHE} branch aim to maximally reduce the computing overhead induced by the bootstrapping operation. Examples of fast bootstrapping schemes are FHEW~\cite{ducas2015fhew}, TFHE~\cite{chillotti2016faster}. 
\textit{Leveled HE} schemes assign levels to different noise sizes and only operate within these noise levels. In this way, they do not have to resort to the 
computationally expensive bootstrapping operation. However, when the noise in a ciphertext exceeds the maximal level, 
they will provide 
incorrect decryptions. Examples of leveled FHE schemes are BGV~\cite{brakerski2014leveled}, BFV~\cite{brakerski2012fully} and CKKS~\cite{cheon2017homomorphic}. Leveled HE schemes can also be extended to FHE, for example, in~\cite{cheon2018bootstrapping}, the authors proposed a bootstrapping procedure to make CKKS an FHE scheme.

In this work, we employ the CKKS scheme. Using CKKS, we can perform computations on vectors of complex values and real values. Suppose we have a plaintext message $\mathbf{z} \in \mathbb{C}^{N/2}$ that we want to encrypt, where $N$ is called polynomial degree modulus and is a power of 2. Before encrypting, $\mathbf{z}$ is encoded into a polynomial $\mathbf{m} \in \mathbb{Z}[X]/\left(X^N+1\right)$. Compared to standard computations on vectors, polynomials provide a better trade-off between security and efficiency. During encoding, $\mathbf{z}$ is also multiplied by a scaling factor $\Delta$ to keep a level of precision. The encoded message $\mathbf{m}$ is then encrypted into $\mathbf{c} \in \left(\mathbb{Z}_q[X]/\left(X^N +1\right)\right)^2$. We can apply addition or multiplication on the ciphertext $\mathbf{c}$ to produce $\mathbf{c}' = f(\mathbf{c})$. In order to get the decrypted message, we need to follow the inverse procedure. First, the ciphertext $\mathbf{c}'$ is decrypted into an encoded message $\mathbf{m}' \in \mathbb{Z}[X]/\left(X^N+1\right)$. Finally, $\mathbf{m}'$ is decoded to produce the transformed plaintext $\mathbf{z}' = f(\mathbf{z})$. In summary, the most important parameters of the CKKS schemes are:
\begin{itemize}
    \item \textbf{Polynomial Modulus (Degree of the polynomial) $N$:} This parameter directly affects the number of coefficients in plaintext polynomial and the size of ciphertext elements. The value $N$ must be a power of 2, such as 1024, 2048, etc\footnote{\url{https://bit.ly/3AB0EfQ}}. The bigger $N$ is, the more secure the scheme. However, 
     the computational overhead incurred is also higher.
    \item \textbf{Coefficient Modulus $\mathcal{C}$:} This is a list of prime numbers that define 
    scheme noise levels. After each multiplication, a different prime in $\mathcal{C}$ is used as the coefficient modulus. After all primes in the list are used, we can no longer 
    perform multiplications, since each homomorphic multiplication on ciphertexts increases the noise level by one.
    \item \textbf{Scaling Factor} $\Delta$: During the encoding process, the plaintext message is multiplied by $\Delta$ to maintain a certain level of precision. $\Delta$ is a positive integer and is often chosen to be a power of 2.
\end{itemize}

\section{Network Architecture}	
\label{sec:network}
In this section, we describe the architecture of our 
1D CNN model. The model has two forms: a local model and a split model. A local model is actually a non-split model in which all the layers of the network are executed by one party. In the split model, the network is split between two parties: the client and the server. 
The first few layers of the network are executed on the client side, while the remaining layers are executed on the server side. We discuss both 
models in more detail 
in~\autoref{subsec:localmodel} and~\autoref{subsec:splitmodel}. Additionally, we discuss the threat model 
considered as well as the 
parties involved in the split model training process, concentrating on their roles and the parameters 
allocated to them throughout the training process.

\subsection{Local Model}
\label{subsec:localmodel}

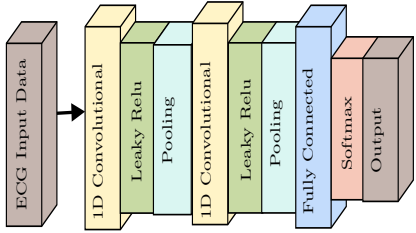
\begin{figure}[!ht]
\centering
 \begin{adjustbox}{width=0.45\textwidth, totalheight=0.25\textwidth}
\tikzset{every picture/.style={line width=0.75pt}} 

\begin{tikzpicture}[x=0.75pt,y=0.75pt,yscale=-1,xscale=1]

\draw  [fill={rgb, 255:red, 255; green, 244; blue, 199 }  ,fill opacity=1 ] (180,154.9) -- (192.9,142) -- (223,142) -- (223,264.1) -- (210.1,277) -- (180,277) -- cycle ; \draw   (223,142) -- (210.1,154.9) -- (180,154.9) ; \draw   (210.1,154.9) -- (210.1,277) ;
\draw  [fill={rgb, 255:red, 200; green, 218; blue, 164 }  ,fill opacity=1 ] (210,164.6) -- (222.6,152) -- (252,152) -- (252,254.5) -- (239.4,267.1) -- (210,267.1) -- cycle ; \draw   (252,152) -- (239.4,164.6) -- (210,164.6) ; \draw   (239.4,164.6) -- (239.4,267.1) ;
\draw  [fill={rgb, 255:red, 218; green, 246; blue, 242 }  ,fill opacity=1 ] (238.2,165.8) -- (252,152) -- (284.2,152) -- (284.2,254.2) -- (270.4,268) -- (238.2,268) -- cycle ; \draw   (284.2,152) -- (270.4,165.8) -- (238.2,165.8) ; \draw   (270.4,165.8) -- (270.4,268) ;
\draw  [fill={rgb, 255:red, 197; green, 181; blue, 175 }  ,fill opacity=1 ] (113,164.9) -- (125.9,152) -- (156,152) -- (156,262.1) -- (143.1,275) -- (113,275) -- cycle ; \draw   (156,152) -- (143.1,164.9) -- (113,164.9) ; \draw   (143.1,164.9) -- (143.1,275) ;

\draw  [fill={rgb, 255:red, 255; green, 244; blue, 199 }  ,fill opacity=1 ] (271.4,152.9) -- (284.3,140) -- (314.4,140) -- (314.4,261.1) -- (301.5,274) -- (271.4,274) -- cycle ; \draw   (314.4,140) -- (301.5,152.9) -- (271.4,152.9) ; \draw   (301.5,152.9) -- (301.5,274) ;
\draw  [fill={rgb, 255:red, 200; green, 218; blue, 164 }  ,fill opacity=1 ] (302,164.6) -- (314.6,152) -- (344,152) -- (344,254.4) -- (331.4,267) -- (302,267) -- cycle ; \draw   (344,152) -- (331.4,164.6) -- (302,164.6) ; \draw   (331.4,164.6) -- (331.4,267) ;
\draw  [fill={rgb, 255:red, 218; green, 246; blue, 242 }  ,fill opacity=1 ] (330.4,164.6) -- (343,152) -- (372.4,152) -- (372.4,254.4) -- (359.8,267) -- (330.4,267) -- cycle ; \draw   (372.4,152) -- (359.8,164.6) -- (330.4,164.6) ; \draw   (359.8,164.6) -- (359.8,267) ;
\draw  [fill={rgb, 255:red, 195; green, 220; blue, 252 }  ,fill opacity=1 ] (359.4,154.9) -- (372.3,142) -- (402.4,142) -- (402.4,263.1) -- (389.5,276) -- (359.4,276) -- cycle ; \draw   (402.4,142) -- (389.5,154.9) -- (359.4,154.9) ; \draw   (389.5,154.9) -- (389.5,276) ;

\draw  [fill={rgb, 255:red, 246; green, 200; blue, 185 }  ,fill opacity=1 ] (389.6,173.9) -- (402.5,161) -- (432.6,161) -- (432.6,246.17) -- (419.7,259.07) -- (389.6,259.07) -- cycle ; \draw   (432.6,161) -- (419.7,173.9) -- (389.6,173.9) ; \draw   (419.7,173.9) -- (419.7,259.07) ;
\draw [line width=1.5]    (156,207) -- (176,206.17) ;
\draw [shift={(180,206)}, rotate = 177.61] [fill={rgb, 255:red, 0; green, 0; blue, 0 }  ][line width=0.08]  [draw opacity=0] (11.61,-5.58) -- (0,0) -- (11.61,5.58) -- cycle    ;
\draw  [fill={rgb, 255:red, 197; green, 181; blue, 175 }  ,fill opacity=1 ] (416.5,173.9) -- (429.4,161) -- (459.5,161) -- (459.5,247.16) -- (446.6,260.06) -- (416.5,260.06) -- cycle ; \draw   (459.5,161) -- (446.6,173.9) -- (416.5,173.9) ; \draw   (446.6,173.9) -- (446.6,260.06) ;

\draw (185.41,273.05) node [anchor=north west][inner sep=0.75pt]  [rotate=-269.64] [align=left] {1D Convolutional};

\draw (277.41,272.05) node [anchor=north west][inner sep=0.75pt]  [rotate=-269.64] [align=left] {1D Convolutional};

\draw (218.41,250.05) node [anchor=north west][inner sep=0.75pt]  [rotate=-269.64] [align=left] {Leaky Relu};

\draw (311.41,249.05) node [anchor=north west][inner sep=0.75pt]  [rotate=-269.64] [align=left] {Leaky Relu};

\draw (244.41,239.05) node [anchor=north west][inner sep=0.75pt]  [rotate=-269.64] [align=left] {Pooling};

\draw (336.41,241.05) node [anchor=north west][inner sep=0.75pt]  [rotate=-269.64] [align=left] {Pooling};

\draw (397.41,242.81) node [anchor=north west][inner sep=0.75pt]  [rotate=-269.64] [align=left] {Softmax};

\draw (422.41,244.26) node [anchor=north west][inner sep=0.75pt]  [rotate=-269.64] [align=left] {Output};

\draw (364.41,269.05) node [anchor=north west][inner sep=0.75pt]  [rotate=-269.64] [align=left] {Fully Connected};

\draw (120.41,267.05) node [anchor=north west][inner sep=0.75pt]  [rotate=-269.64] [align=left] {{\small ECG Input Data}};

\end{tikzpicture}
\end{adjustbox}
	\caption{Local Model}
	\label{fig:localmodel}
\end{figure}

In this section, we describe the non-split version of the 1D CNN model to classify the ECG signals. 
The model was selected from paper~\cite{li2017classification}. 
As can be seen in~\autoref{fig:localmodel}, the non-split version of 1D CNN model contained an input layer, two Conv1D layers, two pooling layers, a fully connected layer and an output layer. The input of each layer is connected to the output of the previous layer. 

\begin{figure}[!ht]
\centering
\begin{adjustbox}{width=0.7\textwidth, totalheight=0.35\textwidth}
\tikzset{every picture/.style={line width=0.75pt}}    

\begin{tikzpicture}[x=0.75pt,y=0.75pt,yscale=-1,xscale=1]
\draw (89,183) node  {\includegraphics[width=52.5pt,height=52.5pt]{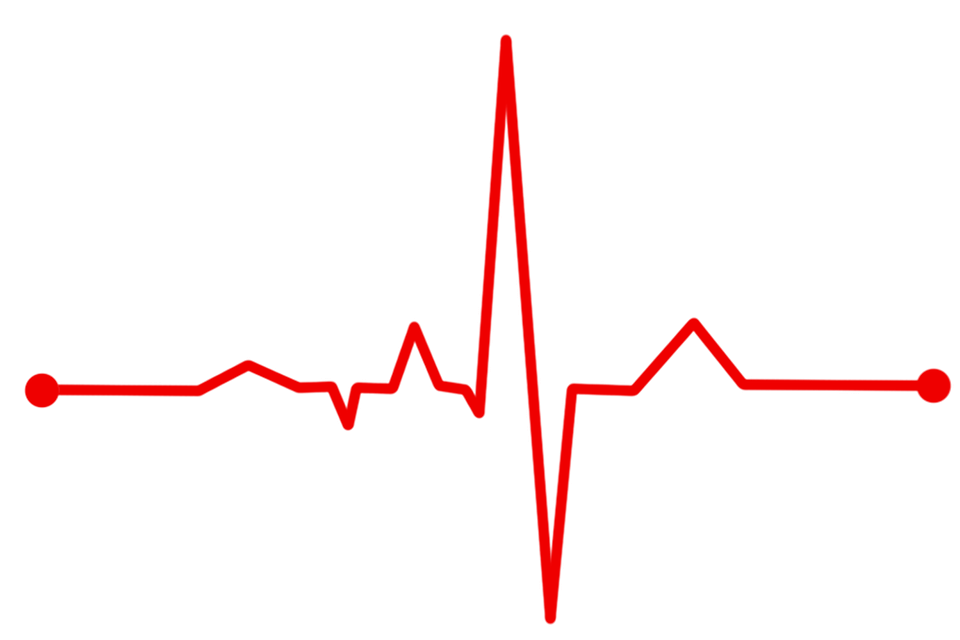}};
\draw   (165,85.2) .. controls (165,82.88) and (166.88,81) .. (169.2,81) -- (181.8,81) .. controls (184.12,81) and (186,82.88) .. (186,85.2) -- (186,275.47) .. controls (186,277.79) and (184.12,279.67) .. (181.8,279.67) -- (169.2,279.67) .. controls (166.88,279.67) and (165,277.79) .. (165,275.47) -- cycle ;
\draw   (171,105.5) .. controls (171,103.01) and (173.01,101) .. (175.5,101) .. controls (177.99,101) and (180,103.01) .. (180,105.5) .. controls (180,107.99) and (177.99,110) .. (175.5,110) .. controls (173.01,110) and (171,107.99) .. (171,105.5) -- cycle ;

\draw   (171,124.5) .. controls (171,122.01) and (173.01,120) .. (175.5,120) .. controls (177.99,120) and (180,122.01) .. (180,124.5) .. controls (180,126.99) and (177.99,129) .. (175.5,129) .. controls (173.01,129) and (171,126.99) .. (171,124.5) -- cycle ;

\draw   (171,174.5) .. controls (171,172.01) and (173.01,170) .. (175.5,170) .. controls (177.99,170) and (180,172.01) .. (180,174.5) .. controls (180,176.99) and (177.99,179) .. (175.5,179) .. controls (173.01,179) and (171,176.99) .. (171,174.5) -- cycle ;

\draw   (171,193.5) .. controls (171,191.01) and (173.01,189) .. (175.5,189) .. controls (177.99,189) and (180,191.01) .. (180,193.5) .. controls (180,195.99) and (177.99,198) .. (175.5,198) .. controls (173.01,198) and (171,195.99) .. (171,193.5) -- cycle ;

\draw   (171,245.5) .. controls (171,243.01) and (173.01,241) .. (175.5,241) .. controls (177.99,241) and (180,243.01) .. (180,245.5) .. controls (180,247.99) and (177.99,250) .. (175.5,250) .. controls (173.01,250) and (171,247.99) .. (171,245.5) -- cycle ;

\draw   (171,264.5) .. controls (171,262.01) and (173.01,260) .. (175.5,260) .. controls (177.99,260) and (180,262.01) .. (180,264.5) .. controls (180,266.99) and (177.99,269) .. (175.5,269) .. controls (173.01,269) and (171,266.99) .. (171,264.5) -- cycle ;

\draw [line width=1.5]  [dash pattern={on 1.69pt off 2.76pt}]  (176,137) -- (176,159) ;

\draw [line width=1.5]  [dash pattern={on 1.69pt off 2.76pt}]  (176,209) -- (176,231) ;

\draw   (373,83.87) .. controls (373,81.55) and (374.88,79.67) .. (377.2,79.67) -- (389.8,79.67) .. controls (392.12,79.67) and (394,81.55) .. (394,83.87) -- (394,276.13) .. controls (394,278.45) and (392.12,280.33) .. (389.8,280.33) -- (377.2,280.33) .. controls (374.88,280.33) and (373,278.45) .. (373,276.13) -- cycle ;
\draw   (379,100.5) .. controls (379,98.01) and (381.01,96) .. (383.5,96) .. controls (385.99,96) and (388,98.01) .. (388,100.5) .. controls (388,102.99) and (385.99,105) .. (383.5,105) .. controls (381.01,105) and (379,102.99) .. (379,100.5) -- cycle ;
\draw   (379,119.5) .. controls (379,117.01) and (381.01,115) .. (383.5,115) .. controls (385.99,115) and (388,117.01) .. (388,119.5) .. controls (388,121.99) and (385.99,124) .. (383.5,124) .. controls (381.01,124) and (379,121.99) .. (379,119.5) -- cycle ;

\draw   (379,169.5) .. controls (379,167.01) and (381.01,165) .. (383.5,165) .. controls (385.99,165) and (388,167.01) .. (388,169.5) .. controls (388,171.99) and (385.99,174) .. (383.5,174) .. controls (381.01,174) and (379,171.99) .. (379,169.5) -- cycle ;

\draw   (379,188.5) .. controls (379,186.01) and (381.01,184) .. (383.5,184) .. controls (385.99,184) and (388,186.01) .. (388,188.5) .. controls (388,190.99) and (385.99,193) .. (383.5,193) .. controls (381.01,193) and (379,190.99) .. (379,188.5) -- cycle ;

\draw   (379,240.5) .. controls (379,238.01) and (381.01,236) .. (383.5,236) .. controls (385.99,236) and (388,238.01) .. (388,240.5) .. controls (388,242.99) and (385.99,245) .. (383.5,245) .. controls (381.01,245) and (379,242.99) .. (379,240.5) -- cycle ;

\draw   (379,259.5) .. controls (379,257.01) and (381.01,255) .. (383.5,255) .. controls (385.99,255) and (388,257.01) .. (388,259.5) .. controls (388,261.99) and (385.99,264) .. (383.5,264) .. controls (381.01,264) and (379,261.99) .. (379,259.5) -- cycle ;

\draw [line width=1.5]  [dash pattern={on 1.69pt off 2.76pt}]  (384,132) -- (384,154) ;

\draw [line width=1.5]  [dash pattern={on 1.69pt off 2.76pt}]  (384,204) -- (384,226) ;

\draw   (212,100.2) .. controls (212,97.88) and (213.88,96) .. (216.2,96) -- (228.8,96) .. controls (231.12,96) and (233,97.88) .. (233,100.2) -- (233,265.8) .. controls (233,268.12) and (231.12,270) .. (228.8,270) -- (216.2,270) .. controls (213.88,270) and (212,268.12) .. (212,265.8) -- cycle ;

\draw   (218,116.98) .. controls (218,114.86) and (220.01,113.13) .. (222.5,113.13) .. controls (224.99,113.13) and (227,114.86) .. (227,116.98) .. controls (227,119.11) and (224.99,120.84) .. (222.5,120.84) .. controls (220.01,120.84) and (218,119.11) .. (218,116.98) -- cycle ;

\draw   (218,133.26) .. controls (218,131.13) and (220.01,129.4) .. (222.5,129.4) .. controls (224.99,129.4) and (227,131.13) .. (227,133.26) .. controls (227,135.39) and (224.99,137.11) .. (222.5,137.11) .. controls (220.01,137.11) and (218,135.39) .. (218,133.26) -- cycle ;

\draw   (218,176.08) .. controls (218,173.96) and (220.01,172.23) .. (222.5,172.23) .. controls (224.99,172.23) and (227,173.96) .. (227,176.08) .. controls (227,178.21) and (224.99,179.94) .. (222.5,179.94) .. controls (220.01,179.94) and (218,178.21) .. (218,176.08) -- cycle ;

\draw   (218,192.36) .. controls (218,190.23) and (220.01,188.5) .. (222.5,188.5) .. controls (224.99,188.5) and (227,190.23) .. (227,192.36) .. controls (227,194.49) and (224.99,196.21) .. (222.5,196.21) .. controls (220.01,196.21) and (218,194.49) .. (218,192.36) -- cycle ;

\draw   (218,236.9) .. controls (218,234.77) and (220.01,233.04) .. (222.5,233.04) .. controls (224.99,233.04) and (227,234.77) .. (227,236.9) .. controls (227,239.03) and (224.99,240.75) .. (222.5,240.75) .. controls (220.01,240.75) and (218,239.03) .. (218,236.9) -- cycle ;

\draw   (218,253.17) .. controls (218,251.04) and (220.01,249.32) .. (222.5,249.32) .. controls (224.99,249.32) and (227,251.04) .. (227,253.17) .. controls (227,255.3) and (224.99,257.03) .. (222.5,257.03) .. controls (220.01,257.03) and (218,255.3) .. (218,253.17) -- cycle ;

\draw [line width=1.5]  [dash pattern={on 1.69pt off 2.76pt}]  (223,143.97) -- (223,162.81) ;

\draw [line width=1.5]  [dash pattern={on 1.69pt off 2.76pt}]  (223,205.63) -- (223,224.48) ;

\draw   (265,115.2) .. controls (265,112.88) and (266.88,111) .. (269.2,111) -- (281.8,111) .. controls (284.12,111) and (286,112.88) .. (286,115.2) -- (286,256.8) .. controls (286,259.12) and (284.12,261) .. (281.8,261) -- (269.2,261) .. controls (266.88,261) and (265,259.12) .. (265,256.8) -- cycle ;

\draw   (271,128.47) .. controls (271,126.7) and (273.01,125.26) .. (275.5,125.26) .. controls (277.99,125.26) and (280,126.7) .. (280,128.47) .. controls (280,130.24) and (277.99,131.68) .. (275.5,131.68) .. controls (273.01,131.68) and (271,130.24) .. (271,128.47) -- cycle ;

\draw   (271,142.02) .. controls (271,140.25) and (273.01,138.81) .. (275.5,138.81) .. controls (277.99,138.81) and (280,140.25) .. (280,142.02) .. controls (280,143.79) and (277.99,145.23) .. (275.5,145.23) .. controls (273.01,145.23) and (271,143.79) .. (271,142.02) -- cycle ;

\draw   (271,177.67) .. controls (271,175.9) and (273.01,174.46) .. (275.5,174.46) .. controls (277.99,174.46) and (280,175.9) .. (280,177.67) .. controls (280,179.44) and (277.99,180.88) .. (275.5,180.88) .. controls (273.01,180.88) and (271,179.44) .. (271,177.67) -- cycle ;

\draw   (271,191.22) .. controls (271,189.45) and (273.01,188.01) .. (275.5,188.01) .. controls (277.99,188.01) and (280,189.45) .. (280,191.22) .. controls (280,192.99) and (277.99,194.43) .. (275.5,194.43) .. controls (273.01,194.43) and (271,192.99) .. (271,191.22) -- cycle ;

\draw   (271,228.3) .. controls (271,226.52) and (273.01,225.09) .. (275.5,225.09) .. controls (277.99,225.09) and (280,226.52) .. (280,228.3) .. controls (280,230.07) and (277.99,231.5) .. (275.5,231.5) .. controls (273.01,231.5) and (271,230.07) .. (271,228.3) -- cycle ;

\draw   (271,241.84) .. controls (271,240.07) and (273.01,238.63) .. (275.5,238.63) .. controls (277.99,238.63) and (280,240.07) .. (280,241.84) .. controls (280,243.62) and (277.99,245.05) .. (275.5,245.05) .. controls (273.01,245.05) and (271,243.62) .. (271,241.84) -- cycle ;

\draw [line width=1.5]  [dash pattern={on 1.69pt off 2.76pt}]  (276,150.93) -- (276,166.62) ;

\draw [line width=1.5]  [dash pattern={on 1.69pt off 2.76pt}]  (276,202.27) -- (276,217.96) ;

\draw (545.5,182.5) node  {\includegraphics[width=45.25pt,height=47.75pt]{images/ECG.png}};

\draw  [color={rgb, 255:red, 128; green, 128; blue, 128 }  ,draw opacity=1 ][dash pattern={on 1.69pt off 2.76pt}][line width=1.5]  (162,92) -- (190,92) -- (190,182) -- (162,182) -- cycle ;

\draw  [color={rgb, 255:red, 128; green, 128; blue, 128 }  ,draw opacity=1 ][dash pattern={on 1.69pt off 2.76pt}][line width=1.5]  (162,114) -- (190,114) -- (190,204) -- (162,204) -- cycle ;
\draw [color={rgb, 255:red, 128; green, 128; blue, 128 }  ,draw opacity=1 ]   (180,105.5) -- (218,116.56) ;

\draw [color={rgb, 255:red, 128; green, 128; blue, 128 }  ,draw opacity=1 ]   (180,174.5) -- (218,116.56) ;

\draw [color={rgb, 255:red, 128; green, 128; blue, 128 }  ,draw opacity=1 ]   (227,116.56) -- (271,127.94) ;

\draw [color={rgb, 255:red, 128; green, 128; blue, 128 }  ,draw opacity=1 ]   (227,132.5) -- (271,127.94) ;

\draw  [color={rgb, 255:red, 128; green, 128; blue, 128 }  ,draw opacity=1 ][dash pattern={on 1.69pt off 2.76pt}][line width=1.5]  (261.33,120.67) -- (289.33,120.67) -- (289.33,185.08) -- (261.33,185.08) -- cycle ;

\draw   (314,129.2) .. controls (314,126.88) and (315.88,125) .. (318.2,125) -- (330.8,125) .. controls (333.12,125) and (335,126.88) .. (335,129.2) -- (335,246.8) .. controls (335,249.12) and (333.12,251) .. (330.8,251) -- (318.2,251) .. controls (315.88,251) and (314,249.12) .. (314,246.8) -- cycle ;

\draw   (320,139.27) .. controls (320,137.83) and (322.01,136.65) .. (324.5,136.65) .. controls (326.99,136.65) and (329,137.83) .. (329,139.27) .. controls (329,140.72) and (326.99,141.9) .. (324.5,141.9) .. controls (322.01,141.9) and (320,140.72) .. (320,139.27) -- cycle ;

\draw   (320,150.34) .. controls (320,148.9) and (322.01,147.72) .. (324.5,147.72) .. controls (326.99,147.72) and (329,148.9) .. (329,150.34) .. controls (329,151.79) and (326.99,152.97) .. (324.5,152.97) .. controls (322.01,152.97) and (320,151.79) .. (320,150.34) -- cycle ;

\draw   (320,179.47) .. controls (320,178.03) and (322.01,176.85) .. (324.5,176.85) .. controls (326.99,176.85) and (329,178.03) .. (329,179.47) .. controls (329,180.92) and (326.99,182.1) .. (324.5,182.1) .. controls (322.01,182.1) and (320,180.92) .. (320,179.47) -- cycle ;
 
\draw   (320,190.54) .. controls (320,189.1) and (322.01,187.92) .. (324.5,187.92) .. controls (326.99,187.92) and (329,189.1) .. (329,190.54) .. controls (329,191.99) and (326.99,193.17) .. (324.5,193.17) .. controls (322.01,193.17) and (320,191.99) .. (320,190.54) -- cycle ;

\draw   (320,220.84) .. controls (320,219.39) and (322.01,218.22) .. (324.5,218.22) .. controls (326.99,218.22) and (329,219.39) .. (329,220.84) .. controls (329,222.29) and (326.99,223.46) .. (324.5,223.46) .. controls (322.01,223.46) and (320,222.29) .. (320,220.84) -- cycle ;

\draw   (320,231.91) .. controls (320,230.46) and (322.01,229.29) .. (324.5,229.29) .. controls (326.99,229.29) and (329,230.46) .. (329,231.91) .. controls (329,233.36) and (326.99,234.53) .. (324.5,234.53) .. controls (322.01,234.53) and (320,233.36) .. (320,231.91) -- cycle ;

\draw [line width=1.5]  [dash pattern={on 1.69pt off 2.76pt}]  (325,157.63) -- (325,170.44) ;

\draw [line width=1.5]  [dash pattern={on 1.69pt off 2.76pt}]  (325,199.57) -- (325,212.39) ;

\draw [color={rgb, 255:red, 128; green, 128; blue, 128 }  ,draw opacity=1 ]   (280,127.94) -- (320,139.27) ;

\draw [color={rgb, 255:red, 128; green, 128; blue, 128 }  ,draw opacity=1 ]   (278.67,175.67) -- (320,139.27) ;

\draw    (338,190) -- (367.5,190) ;
\draw [shift={(370.5,190)}, rotate = 180] [fill={rgb, 255:red, 0; green, 0; blue, 0 }  ][line width=0.08]  [draw opacity=0] (8.93,-4.29) -- (0,0) -- (8.93,4.29) -- cycle    ;

\draw    (125,190) -- (154.5,190) ;
\draw [shift={(157.5,190)}, rotate = 180] [fill={rgb, 255:red, 0; green, 0; blue, 0 }  ][line width=0.08]  [draw opacity=0] (8.93,-4.29) -- (0,0) -- (8.93,4.29) -- cycle    ;
\draw   (173.33,336) .. controls (173.33,340.67) and (175.66,343) .. (180.33,343) -- (244.79,343) .. controls (251.46,343) and (254.79,345.33) .. (254.79,350) .. controls (254.79,345.33) and (258.12,343) .. (264.79,343)(261.79,343) -- (325.33,343) .. controls (330,343) and (332.33,340.67) .. (332.33,336) ;
\draw   (421,132.2) .. controls (421,129.88) and (422.88,128) .. (425.2,128) -- (437.8,128) .. controls (440.12,128) and (442,129.88) .. (442,132.2) -- (442,257.8) .. controls (442,260.12) and (440.12,262) .. (437.8,262) -- (425.2,262) .. controls (422.88,262) and (421,260.12) .. (421,257.8) -- cycle ;

\draw   (427,142.27) .. controls (427,140.83) and (429.01,139.65) .. (431.5,139.65) .. controls (433.99,139.65) and (436,140.83) .. (436,142.27) .. controls (436,143.72) and (433.99,144.9) .. (431.5,144.9) .. controls (429.01,144.9) and (427,143.72) .. (427,142.27) -- cycle ;

\draw   (427,153.34) .. controls (427,151.9) and (429.01,150.72) .. (431.5,150.72) .. controls (433.99,150.72) and (436,151.9) .. (436,153.34) .. controls (436,154.79) and (433.99,155.97) .. (431.5,155.97) .. controls (429.01,155.97) and (427,154.79) .. (427,153.34) -- cycle ;

\draw   (427,223.84) .. controls (427,222.39) and (429.01,221.22) .. (431.5,221.22) .. controls (433.99,221.22) and (436,222.39) .. (436,223.84) .. controls (436,225.29) and (433.99,226.46) .. (431.5,226.46) .. controls (429.01,226.46) and (427,225.29) .. (427,223.84) -- cycle ;

\draw   (427,234.91) .. controls (427,233.46) and (429.01,232.29) .. (431.5,232.29) .. controls (433.99,232.29) and (436,233.46) .. (436,234.91) .. controls (436,236.36) and (433.99,237.53) .. (431.5,237.53) .. controls (429.01,237.53) and (427,236.36) .. (427,234.91) -- cycle ;

\draw [line width=1.5]  [dash pattern={on 1.69pt off 2.76pt}]  (432,160.63) -- (432,173.44) ;

\draw [line width=1.5]  [dash pattern={on 1.69pt off 2.76pt}]  (432,202.57) -- (432,215.39) ;

\draw   (473,154.2) .. controls (473,151.88) and (474.88,150) .. (477.2,150) -- (489.8,150) .. controls (492.12,150) and (494,151.88) .. (494,154.2) -- (494,237.8) .. controls (494,240.12) and (492.12,242) .. (489.8,242) -- (477.2,242) .. controls (474.88,242) and (473,240.12) .. (473,237.8) -- cycle ;

\draw   (478.5,226.5) .. controls (478.5,224.01) and (480.51,222) .. (483,222) .. controls (485.49,222) and (487.5,224.01) .. (487.5,226.5) .. controls (487.5,228.99) and (485.49,231) .. (483,231) .. controls (480.51,231) and (478.5,228.99) .. (478.5,226.5) -- cycle ;

\draw [color={rgb, 255:red, 128; green, 128; blue, 128 }  ,draw opacity=1 ]   (436,142.27) -- (478.5,162.5) ;

\draw [line width=1.5]  [dash pattern={on 1.69pt off 2.76pt}]  (431.5,182.18) -- (431.5,195) ;

\draw [color={rgb, 255:red, 128; green, 128; blue, 128 }  ,draw opacity=1 ]   (436,153.34) -- (478.5,226.5) ;
 
\draw   (478.5,162.5) .. controls (478.5,160.01) and (480.51,158) .. (483,158) .. controls (485.49,158) and (487.5,160.01) .. (487.5,162.5) .. controls (487.5,164.99) and (485.49,167) .. (483,167) .. controls (480.51,167) and (478.5,164.99) .. (478.5,162.5) -- cycle ;

\draw [color={rgb, 255:red, 128; green, 128; blue, 128 }  ,draw opacity=1 ]   (436,142.27) -- (463.67,194.67) -- (478.5,226.5) ;

\draw [color={rgb, 255:red, 128; green, 128; blue, 128 }  ,draw opacity=1 ]   (436,153.34) -- (478.5,162.5) ;

\draw [color={rgb, 255:red, 128; green, 128; blue, 128 }  ,draw opacity=1 ]   (436,223.84) -- (478.5,162.5) ;
 
\draw [color={rgb, 255:red, 128; green, 128; blue, 128 }  ,draw opacity=1 ]   (436,234.91) -- (478.5,162.5) ;

\draw [color={rgb, 255:red, 128; green, 128; blue, 128 }  ,draw opacity=1 ]   (436,223.84) -- (478.5,226.5) ;
 
\draw [color={rgb, 255:red, 128; green, 128; blue, 128 }  ,draw opacity=1 ]   (436,234.91) -- (478.5,226.5) ;

\draw [line width=1.5]  [dash pattern={on 1.69pt off 2.76pt}]  (482.83,173.85) -- (482.83,182.67) -- (482.83,186.67) ;

\draw [color={rgb, 255:red, 128; green, 128; blue, 128 }  ,draw opacity=1 ]   (388,100.5) -- (427,142.27) ;

\draw [color={rgb, 255:red, 128; green, 128; blue, 128 }  ,draw opacity=1 ]   (388,100.5) -- (427,153.34) ;
 
\draw [color={rgb, 255:red, 128; green, 128; blue, 128 }  ,draw opacity=1 ]   (388,119.5) -- (427,142.27) ;

\draw [color={rgb, 255:red, 128; green, 128; blue, 128 }  ,draw opacity=1 ]   (388,119.5) -- (427,153.34) ;

\draw [color={rgb, 255:red, 128; green, 128; blue, 128 }  ,draw opacity=1 ]   (388,100.5) -- (427,223.84) ;

\draw [color={rgb, 255:red, 128; green, 128; blue, 128 }  ,draw opacity=1 ]   (388,119.5) -- (427,223.84) ;

\draw [color={rgb, 255:red, 128; green, 128; blue, 128 }  ,draw opacity=1 ]   (388,169.5) -- (427,142.27) ;

\draw [color={rgb, 255:red, 128; green, 128; blue, 128 }  ,draw opacity=1 ]   (388,169.5) -- (427,223.84) ;

\draw [color={rgb, 255:red, 128; green, 128; blue, 128 }  ,draw opacity=1 ]   (388,169.5) -- (427,153.34) ;

\draw [color={rgb, 255:red, 128; green, 128; blue, 128 }  ,draw opacity=1 ]   (388,188.5) -- (427,142.27) ;

\draw [color={rgb, 255:red, 128; green, 128; blue, 128 }  ,draw opacity=1 ]   (388,188.5) -- (427,153.34) ;

\draw [color={rgb, 255:red, 128; green, 128; blue, 128 }  ,draw opacity=1 ]   (388,188.5) -- (427,223.84) ;

\draw [color={rgb, 255:red, 128; green, 128; blue, 128 }  ,draw opacity=1 ]   (388,188.5) -- (427,234.91) ;

\draw [color={rgb, 255:red, 128; green, 128; blue, 128 }  ,draw opacity=1 ]   (388,240.5) -- (427,234.91) ;

\draw [color={rgb, 255:red, 128; green, 128; blue, 128 }  ,draw opacity=1 ]   (388,257.33) -- (427,234.91) ;

\draw [color={rgb, 255:red, 128; green, 128; blue, 128 }  ,draw opacity=1 ]   (388,257.33) -- (427,223.84) ;

\draw [color={rgb, 255:red, 128; green, 128; blue, 128 }  ,draw opacity=1 ]   (388,240.5) -- (427,223.84) ;

\draw [color={rgb, 255:red, 128; green, 128; blue, 128 }  ,draw opacity=1 ]   (388,240.5) -- (427,153.34) ;

\draw [color={rgb, 255:red, 128; green, 128; blue, 128 }  ,draw opacity=1 ]   (388,240.5) -- (427,142.27) ;

\draw [color={rgb, 255:red, 128; green, 128; blue, 128 }  ,draw opacity=1 ]   (388,257.33) -- (427,153.34) ;

\draw [color={rgb, 255:red, 128; green, 128; blue, 128 }  ,draw opacity=1 ]   (388,257.33) -- (427,142.27) ;

\draw [color={rgb, 255:red, 128; green, 128; blue, 128 }  ,draw opacity=1 ]   (388,100.5) -- (427,234.91) ;

\draw [color={rgb, 255:red, 128; green, 128; blue, 128 }  ,draw opacity=1 ]   (388,119.5) -- (427,234.91) ;
 
\draw [line width=1.5]  [dash pattern={on 1.69pt off 2.76pt}]  (483.17,192.33) -- (483.17,201.15) -- (483.17,205.15) ;
 
\draw [line width=1.5]  [dash pattern={on 1.69pt off 2.76pt}]  (483.17,205.15) -- (483.17,213.97) -- (483.17,217.97) ;
 
\draw   (381,289) .. controls (381,293.67) and (383.33,296) .. (388,296) -- (425.47,296) .. controls (432.14,296) and (435.47,298.33) .. (435.47,303) .. controls (435.47,298.33) and (438.8,296) .. (445.47,296)(442.47,296) -- (480.67,296) .. controls (485.34,296) and (487.67,293.67) .. (487.67,289) ;

\draw    (496.33,189.67) -- (510,189.67) ;
\draw [shift={(513,189.67)}, rotate = 180] [fill={rgb, 255:red, 0; green, 0; blue, 0 }  ][line width=0.08]  [draw opacity=0] (8.93,-4.29) -- (0,0) -- (8.93,4.29) -- cycle    ;

\draw (130,280) node [anchor=north west][inner sep=0.75pt]   [align=left] {{\footnotesize \textbf{\ \ \ \ \ \ Conv1D}}\\{\footnotesize \textbf{ Feature maps}}\\{\footnotesize \textbf{ \ \ \ 7@1x16}}};

\draw (393.67,267.8) node [anchor=north west][inner sep=0.75pt]   [align=left] {{\footnotesize \textbf{Fully connected}}\\{\footnotesize \textbf{	\ \ \ \ layer }}};

\draw (67.33,124.47) node [anchor=north west][inner sep=0.75pt]   [align=left] {{\footnotesize \textbf{ \ Input}}\\{\footnotesize \textbf{1x128}}};

\draw (183.33,47.8) node [anchor=north west][inner sep=0.75pt]   [align=left] {{\footnotesize \textbf{Pooling layer 1}}\\{\footnotesize \textbf{Feature maps}}\\{\footnotesize \textbf{ \ \ \ Max@2}}};

\draw (284,58.47) node [anchor=north west][inner sep=0.75pt]   [align=left] {{\footnotesize \textbf{Pooling layer 2}}\\{\footnotesize \textbf{Feature maps}}\\{\footnotesize \textbf{ \ \ \ Max@2}}};

\draw (239,279.67) node [anchor=north west][inner sep=0.75pt]   [align=left] {{\footnotesize \textbf{\ \ \ \ \  Conv1D}}\\{\footnotesize \textbf{ Feature maps}}\\{\footnotesize \textbf{ \ \ \ 5@16x8}}};

\draw (394.67,303.8) node [anchor=north west][inner sep=0.75pt]   [align=left] {{\footnotesize \textbf{Classification }}};

\draw (188.67,350.67) node [anchor=north west][inner sep=0.75pt]   [align=left] {{\footnotesize \textbf{Feature extraction}}};

\draw (460.33,129.13) node [anchor=north west][inner sep=0.75pt]   [align=left] {{\footnotesize \textbf{Softmax}}};

\draw (515,125.13) node [anchor=north west][inner sep=0.75pt]   [align=left] {{\footnotesize \textbf{ \ Output}}\\{\footnotesize \textbf{\ \ \ 1x5}}};
\end{tikzpicture}
\end{adjustbox}
	\caption{Architecture and parameters of the CNN model used for ECG classification}
	\label{fig:classification}
\end{figure}

As shown in~\autoref{fig:classification}, the 1D CNN model has two parts. The first one is feature extraction, which consists of two types of layers: Conv1D and pooling layer. 
This latter automatically learns the features from the raw input data. In the Conv1D layer, the 1D CNN model carries out the convolution operation on the input to generate the corresponding one-dimensional feature maps. Different convolution kernels are used to extract 
varying features from the input signal as convolution kernels of size 7 are used for the first Conv1D layer, while kernels of size 5 are used for the second Conv1D layer. Consequently, the Conv1D serves as a filter 
that learns the necessary characteristics of the original signal. The outputs of the Conv1D layers are called feature maps, or activation maps. 
Following, each feature map is now processed using the max pooling operation. The pooling layer reduces the dimension of the feature maps and prevents the network from over-fitting. The second part of the neural network is classification. 
Through it, signals are classified accurately by utilizing the extracted features. It consists of a fully connected layer and a softmax layer. The fully connected layer performs a weighted sum of the inputs and adds bias. It integrates and normalizes highly abstracted features 
used as the input of a softmax classifier to classify the segments into different types. More detailed descriptions of the hyper-parameters for each layer are given in~\autoref{fig:classification}.

\subsection{Split Model}
\label{subsec:splitmodel}

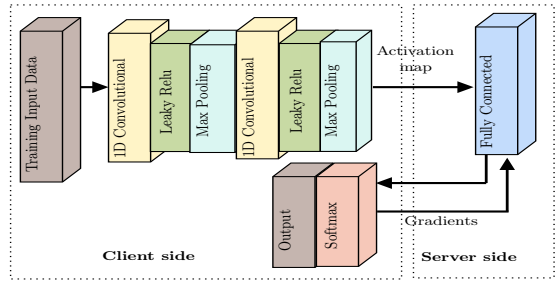
\begin{figure}[!ht]
\centering
\begin{adjustbox}{width=0.6\textwidth, totalheight=0.3\textwidth}
\tikzset{every picture/.style={line width=0.75pt}} 

\begin{tikzpicture}[x=0.75pt,y=0.75pt,yscale=-1,xscale=1]
\draw  [fill={rgb, 255:red, 255; green, 244; blue, 199 }  ,fill opacity=1 ] (167,155.9) -- (179.9,143) -- (210,143) -- (210,265.1) -- (197.1,278) -- (167,278) -- cycle ; \draw   (210,143) -- (197.1,155.9) -- (167,155.9) ; \draw   (197.1,155.9) -- (197.1,278) ;
\draw  [fill={rgb, 255:red, 200; green, 218; blue, 164 }  ,fill opacity=1 ] (197,165.6) -- (209.6,153) -- (239,153) -- (239,255.5) -- (226.4,268.1) -- (197,268.1) -- cycle ; \draw   (239,153) -- (226.4,165.6) -- (197,165.6) ; \draw   (226.4,165.6) -- (226.4,268.1) ;
\draw  [fill={rgb, 255:red, 218; green, 246; blue, 242 }  ,fill opacity=1 ] (225.2,166.8) -- (239,153) -- (271.2,153) -- (271.2,255.2) -- (257.4,269) -- (225.2,269) -- cycle ; \draw   (271.2,153) -- (257.4,166.8) -- (225.2,166.8) ; \draw   (257.4,166.8) -- (257.4,269) ;
\draw  [fill={rgb, 255:red, 197; green, 181; blue, 175 }  ,fill opacity=1 ] (104,153.9) -- (116.9,141) -- (147,141) -- (147,284.1) -- (134.1,297) -- (104,297) -- cycle ; \draw   (147,141) -- (134.1,153.9) -- (104,153.9) ; \draw   (134.1,153.9) -- (134.1,297) ;
\draw  [fill={rgb, 255:red, 255; green, 244; blue, 199 }  ,fill opacity=1 ] (258.4,153.9) -- (271.3,141) -- (301.4,141) -- (301.4,262.1) -- (288.5,275) -- (258.4,275) -- cycle ; \draw   (301.4,141) -- (288.5,153.9) -- (258.4,153.9) ; \draw   (288.5,153.9) -- (288.5,275) ;
\draw  [fill={rgb, 255:red, 200; green, 218; blue, 164 }  ,fill opacity=1 ] (289,165.6) -- (301.6,153) -- (331,153) -- (331,255.4) -- (318.4,268) -- (289,268) -- cycle ; \draw   (331,153) -- (318.4,165.6) -- (289,165.6) ; \draw   (318.4,165.6) -- (318.4,268) ;
\draw  [fill={rgb, 255:red, 218; green, 246; blue, 242 }  ,fill opacity=1 ] (318.4,163.98) -- (329.38,153) -- (355,153) -- (355,257.02) -- (344.02,268) -- (318.4,268) -- cycle ; \draw   (355,153) -- (344.02,163.98) -- (318.4,163.98) ; \draw   (344.02,163.98) -- (344.02,268) ;
\draw  [fill={rgb, 255:red, 195; green, 220; blue, 252 }  ,fill opacity=1 ] (429.4,149.9) -- (442.3,137) -- (472.4,137) -- (472.4,258.1) -- (459.5,271) -- (429.4,271) -- cycle ; \draw   (472.4,137) -- (459.5,149.9) -- (429.4,149.9) ; \draw   (459.5,149.9) -- (459.5,271) ;
\draw  [fill={rgb, 255:red, 197; green, 181; blue, 175 }  ,fill opacity=1 ] (284.5,291.9) -- (297.4,279) -- (327.5,279) -- (327.5,365.16) -- (314.6,378.06) -- (284.5,378.06) -- cycle ; \draw   (327.5,279) -- (314.6,291.9) -- (284.5,291.9) ; \draw   (314.6,291.9) -- (314.6,378.06) ;
\draw  [fill={rgb, 255:red, 246; green, 200; blue, 185 }  ,fill opacity=1 ] (315.6,291.9) -- (328.5,279) -- (358.6,279) -- (358.6,364.17) -- (345.7,377.07) -- (315.6,377.07) -- cycle ; \draw   (358.6,279) -- (345.7,291.9) -- (315.6,291.9) ; \draw   (345.7,291.9) -- (345.7,377.07) ;
\draw [line width=1.5]    (147,207) -- (163,207) ;
\draw [shift={(167,207)}, rotate = 180] [fill={rgb, 255:red, 0; green, 0; blue, 0 }  ][line width=0.08]  [draw opacity=0] (11.61,-5.58) -- (0,0) -- (11.61,5.58) -- cycle    ;
\draw [line width=1.5]    (356,207) -- (424,207) ;
\draw [shift={(428,207)}, rotate = 180] [fill={rgb, 255:red, 0; green, 0; blue, 0 }  ][line width=0.08]  [draw opacity=0] (11.61,-5.58) -- (0,0) -- (11.61,5.58) -- cycle    ;
\draw [line width=1.5]    (437,298) -- (362,298) ;
\draw [shift={(358,298)}, rotate = 360] [fill={rgb, 255:red, 0; green, 0; blue, 0 }  ][line width=0.08]  [draw opacity=0] (11.61,-5.58) -- (0,0) -- (11.61,5.58) -- cycle    ;
\draw  [dash pattern={on 0.84pt off 2.51pt}] (97,129) -- (376.5,129) -- (376.5,389) -- (97,389) -- cycle ;
\draw  [dash pattern={on 0.84pt off 2.51pt}] (385,129) -- (485.5,129) -- (485.5,388) -- (385,388) -- cycle ;
\draw [line width=1.5]    (437,271) -- (437,298) ;
\draw [line width=1.5]    (359,325) -- (453,325) ;
\draw [line width=1.5]    (452,325) -- (452,278) ;
\draw [shift={(452,274)}, rotate = 90] [fill={rgb, 255:red, 0; green, 0; blue, 0 }  ][line width=0.08]  [draw opacity=0] (11.61,-5.58) -- (0,0) -- (11.61,5.58) -- cycle    ;
\draw (169.42,276.07) node [anchor=north west][inner sep=0.75pt]  [rotate=-269.64] [align=left] {1D Convolutional};
\draw (261.42,275.07) node [anchor=north west][inner sep=0.75pt]  [rotate=-269.64] [align=left] {1D Convolutional};
\draw (202.42,253.07) node [anchor=north west][inner sep=0.75pt]  [rotate=-269.64] [align=left] {Leaky Relu};
\draw (295.42,252.07) node [anchor=north west][inner sep=0.75pt]  [rotate=-269.64] [align=left] {Leaky Relu};
\draw (228.42,256.07) node [anchor=north west][inner sep=0.75pt]  [rotate=-269.64] [align=left] {Max Pooling};
\draw (320.42,256.07) node [anchor=north west][inner sep=0.75pt]  [rotate=-269.64] [align=left] {Max Pooling};
\draw (161,361) node [anchor=north west][inner sep=0.75pt]   [align=left] {\textbf{Client side}};
\draw (389,362) node [anchor=north west][inner sep=0.75pt]   [align=left] {\textbf{Server side}};
\draw (320.42,361.83) node [anchor=north west][inner sep=0.75pt]  [rotate=-269.64] [align=left] {Softmax};
\draw (287.42,364.28) node [anchor=north west][inner sep=0.75pt]  [rotate=-269.64] [align=left] {Output};
\draw (106.42,289.07) node [anchor=north west][inner sep=0.75pt]  [rotate=-269.64] [align=left] {Training Input Data};
\draw (357,167) node [anchor=north west][inner sep=0.75pt]   [align=left] {Activation\\ \ \ \ \ map};
\draw (377,328) node [anchor=north west][inner sep=0.75pt]   [align=left] {Gradients};
\draw (431.42,266.07) node [anchor=north west][inner sep=0.75pt]  [rotate=-269.64] [align=left] {Fully Connected};

\end{tikzpicture}

\end{adjustbox}
	\caption{U-shaped Split Model}
	\label{fig:splitmodel}
\end{figure}

In this section, we discuss the split version of the above 1D CNN model.  We split the 1D CNN model into multiple parts, each of which is processed and computed by a separate party. For our setting, we consider two parties, a client and a server, with one part of the model executed on the client and another part of the model executed on the server. Both parties participate in training the 1D CNN model. As can be seen in~\autoref{fig:splitmodel}, activation maps and gradients are passed between client and server to collaboratively train the joint model, but neither can access the other's data. In addition, we used the U-shaped SL, where the first few layers $(1 \dots l)$ and the last layer $L$ are executed on the client side, while the remaining layers $(L-1)$ are executed on the server side. As a result, only the client has access to the model's output, while the server does not know the 1D CNN model's output. We will not go into the details of protocol training 
with the plaintext activation map as the authors in~\cite{khan2023split} have shown that SL can be applied into 1D CNN without the model classification accuracy degradation. Note that with the current split model architecture, we only have one fully connected layer on the server side, which goes against the advantage of SL where the client can outsource a big part of the neural network to the server to reduce computation cost on the client side. The reason for this is that in our encrypted training protocol, we will encrypt the activation maps with HE before sending them to the server (more details in~\autoref{sec:splitmodel}). With the current limitations of HE, e.g. big computation complexity and difficulties in non-linear calculations\ldots, in this work, we limit to have only one linear layer on the server side. Adding more layers on the server side will be done in future works. 

\subsection{Threat Model}
\label{subsec:threatmodel}

For this setting, we consider a semi-honest threat model in which both parties follow the protocol while trying to gather as much information as possible from the received messages. As mentioned earlier, in the SL context, we have two parties: a client and a server, each of which plays a specific role and has access to specified parameters. Both parties do not access each other's devices and cannot target attacks on each other. More specifically, the server performs all its operations as defined, does not collude with the client but is curious about the raw data stored with the client. The server's goal is to reconstruct the raw data from the activation maps of the split layer delivered from the client during the forward propagation and also the gradients delivered by the client during backward propagation. Furthermore, we assume that the clients are trustworthy and that they will participate in the learning process as long as the raw data remains in their possession, however, the client also tries to learn the server's neural network weights and biases. In short, both client and server follow the protocol but may try to extract maximum information from the exchanged messages.

\section{Privacy Leakage Analysis}
\label{sec:privacyLeakageAnal}
In this section, we report the privacy leakage we 
noted in the backward propagation phase, while training the U-shaped split learning model on HE encrypted data proposed in~\cite{khan2023split}. Here, we also present 
possible solutions to mitigate this privacy leakage and in~\autoref{sec:splitmodel} we discuss in detail how we used homomorphic encryption to address this privacy leakage. In addition, we provide an evaluation of the privacy leakage of the Algorithm~3 in~\cite{khan2023split}. 


In SL, the training is performed through a vertically distributed back-propagation that requires clients to exclusively share 
the intermediate network's output; rather than the raw, private training instances~\cite{pasquini2021unleashing}. In~\cite{khan2023split}, before the training phase, the client decides on the model's architecture and hyper-parameters and sends the required information 
to the server. Hence, the authors assumed the server has no information on the client side architecture and its weights. This is one of the important security features of SL: 
it conceals information about the model's architecture and hyper-parameters. Furthermore, because the client encrypts the activation map before transmitting it to the server during forward propagation, the next assumption is that the server is unaware of the activation maps as they are encrypted, as long as the HE algorithm is secure. However, during 
backward propagation in the Algorithm 3 of~\cite{khan2023split}, the client sends both the gradients of error $J$ w.r.t $a^{L}$ ($\frac{\partial J}{\partial \mathbf{a}^{(L)}}$) and $w^{L}$ ($\frac{\partial J}{\partial \boldsymbol{w}^{(L)}}$) to the server, allowing the server to determine the activation map. 
Following, we present detailed information 
on the privacy breach we detected during 
backpropagation.

In the backward pass in Algorithm 3 from~\cite{khan2023split}, the client calculates $\frac{\partial J}{\partial \mathbf{a}^{(L)}}$ and $\frac{\partial J}{\partial \boldsymbol{w}^{(L)}}$ and sends them in plaintext to server to continue backward propagation, where $\frac{\partial J}{\partial \boldsymbol{w}^{(L)}}$ is calculated by the client using the chain rule as follows:
\begin{equation}\label{eq:dJdw}
	\frac{\partial J}{\partial \boldsymbol{w}^{(L)}} = \frac{\partial J}{\partial \mathbf{a}^{(L)}} \frac{\partial \mathbf{a}^{(L)}}{\partial \boldsymbol{w}^{(L)}}.
\end{equation}
Furthermore, since on the server side there is only one linear layer, 
we have:
\begin{equation}
	\frac{\partial \mathbf{a}^{(L)}}{\partial \boldsymbol{w}^{(L)}} = \mathbf{a}^{(l)}.
\end{equation}
Equation~\eqref{eq:dJdw} now becomes:
\begin{equation}\label{eq:dJdw(2)}
	\frac{\partial J}{\partial \boldsymbol{w}^{(L)}} = \frac{\partial J}{\partial \mathbf{a}^{(L)}} \cdot \mathbf{a}^{(l)}.
\end{equation}
Therefore, if the client sends both $\frac{\partial J}{\partial \boldsymbol{w}^{(L)}}$ and $\frac{\partial J}{\partial \mathbf{a}^{(L)}}$ to the server in plaintext, the server can then find out the plaintext activation maps by multiplying both sides of equation~\eqref{eq:dJdw(2)} with $\left( \frac{\partial J}{\partial \mathbf{a}^{(L)}} \right)^{-1}$ to figure out $\mathbf{a}^{(l)}$ as:
\begin{align}\label{eq:privacyLeakage}
\begin{split}
	\left( \frac{\partial J}{\partial \mathbf{a}^{(L)}} \right)^{-1} \cdot \frac{\partial J}{\partial \boldsymbol{w}^{(L)}} 
	= \left( \frac{\partial J}{\partial \mathbf{a}^{(L)}} \right)^{-1} \cdot \left( \frac{\partial J}{\partial \mathbf{a}^{(L)}} \right) \cdot \mathbf{a}^{(l)} 
	= \mathbf{a}^{(l)} \text{.}
\end{split}
\end{align}
This privacy leakage will happen if the server is able to find the inverse matrix of $\left( \frac{\partial J}{\partial \mathbf{a}^{(L)}} \right)^{-1}$, which is highly possible as the server can manipulate the training batch size to make $\mathbf{a}^{(L)}$ and $\left( \frac{\partial J}{\partial \mathbf{a}^{(L)}} \right)$ square matrices. In the next section, we provide a more detailed evaluation of this privacy leakage analysis. 
To resolve this privacy leakage, we propose a new training protocol in~\autoref{sec:splitmodel}.


\subsection{Privacy Leakage Evaluation}
\label{subsec:privLeakEval}
Given that the server instructs the client to train with the proper batch size to make $\frac{\partial J}{\partial \mathbf{a}^{(L)}}$ a square matrix, after the server receives the gradients that include $\frac{\partial J}{\partial \mathbf{a}^{(L)}}$ and $\frac{\partial J}{\partial \boldsymbol{w}^{(L)}}$ from the client, the server tries to reconstruct $\mathbf{a}^{(l)}$ like in~\autoref{eq:privacyLeakage}. We evaluate this privacy leakage by building the attack on top of the code provided by the authors of~\cite{khan2023split}. 

\begin{figure}[!h]
	\centering
	\includegraphics[width=0.65\textwidth, totalheight=0.25\textwidth]{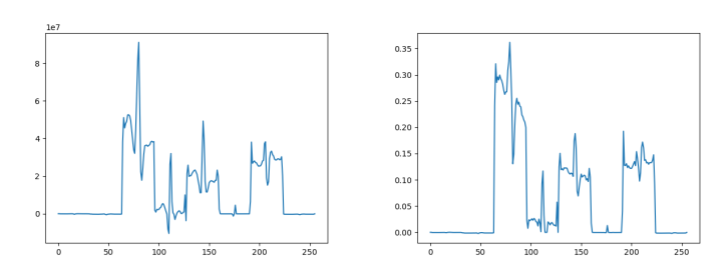}
	\caption{Server's reconstructed (Left) and Client's plaintext (Right) activation maps}
	\label{fig:a_inv}
\end{figure}

\begin{figure}[!h]
	\centering
	\includegraphics[width=0.65\textwidth, totalheight=0.3\textwidth]{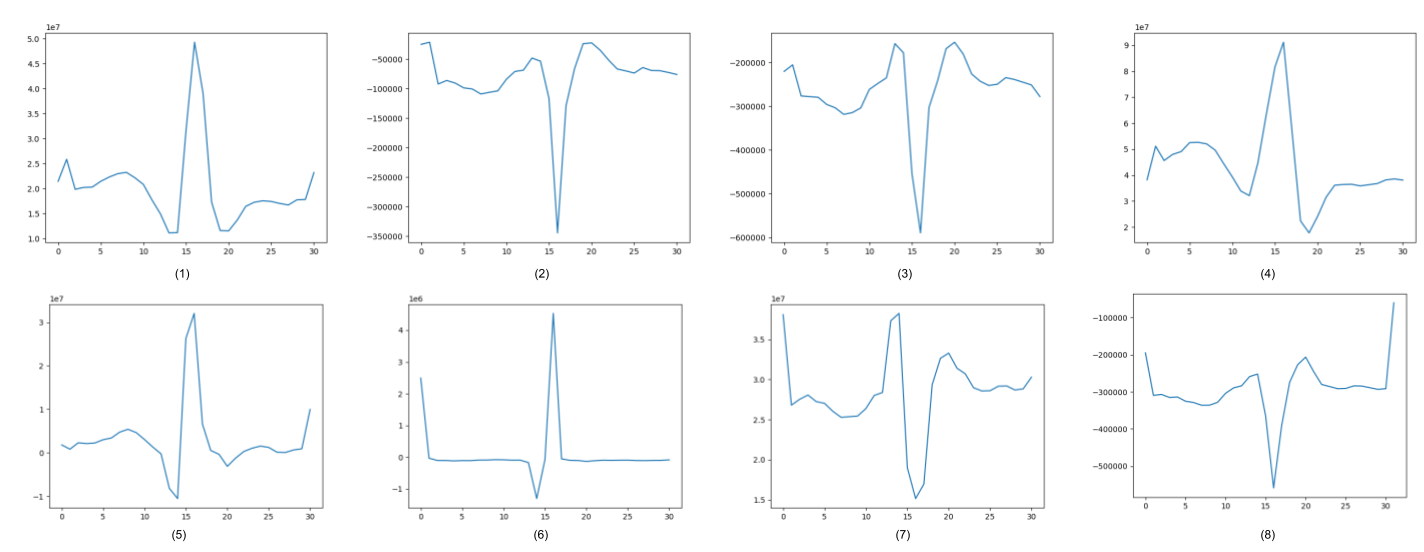}
	\caption{The reconstructed activation map in chunks.}
	\label{fig:a_inv_chunks}
\end{figure}

\begin{figure}[!h]
	\centering
	\includegraphics[width=0.65\textwidth, totalheight=0.25\textwidth]{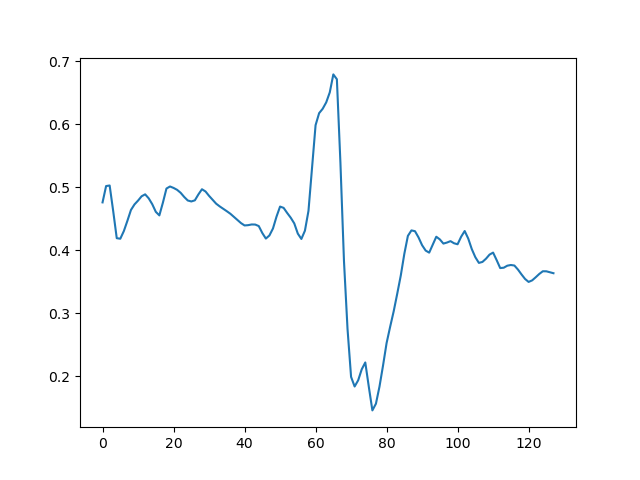}
	\caption{The client's plaintext data.}
	\label{fig:plaintext_input}
\end{figure}

First, the server instructs the client to train with the batch size of 5 (since we have 5 output classes), making $\frac{\partial J}{\partial \mathbf{a}^{(L)}}$ a square matrix. Then, using~\autoref{eq:privacyLeakage}, the server is able to reconstruct an activation map plotted in~\autoref{fig:a_inv} (left), which looks very similar compared to the client's plaintext activation map~\autoref{fig:a_inv} (right). 
Furthermore, when the server plots the reconstructed activation map in chunks of 32 data points as in~\autoref{fig:a_inv_chunks}, it can reveal some very similar patterns compared to the client's original data plotted in~\autoref{fig:plaintext_input}. Observing two figures, we can see that the 4th, 5th and 7th chunk look very similar to the plaintext data, which shows the effectiveness of the privacy leakage we presented in~\autoref{sec:privacyLeakageAnal}. Furthermore, note that in~\cite{khan2023split}, the neural network can only produce about 88\% accuracy on the test dataset. Once the neural network gets to higher accuracy (e.g. 99\%), the privacy leakage can become even more evident.

\section{Split Model Training Protocol using Encrypted Activation Map}
\label{sec:splitmodel}

In this section, we discuss in detail the U-shaped split model with the encrypted activation map. We refer to this protocol as HESplit network. Before delving into the details, consider a model with $L$ layers in total, with the $L^{th}$ layer serving as the model's output layer. Assume the model is divided into layers $l$ and $l+1$. As discussed earlier, most of the layers of the 1D CNN model are executed on the client side, while only one layer is executed on the server side. Hence, the client owns the first $l$ layers from layer~1 to layer $l+1$, as well as the final output layer, whilst the server owns only one linear layer. The main reason for having only one linear layer on the server side is due to computational constraints when training on HE encrypted data. Furthermore, 
more layers on the server's side require us to insert non-linear operations, which have shown difficulties for encrypted computation. In particular, for the CKKS scheme, we need to resort to approximation~\cite{khan2021blind} which may lead to further 
degradation in accuracy. Our future works will focus on 
additional layers on the server side for encrypted split training. 
Below, we describe the U-shaped SL training protocol with the encrypted activation map.


\subsubsection{Socket initialization and context generation}

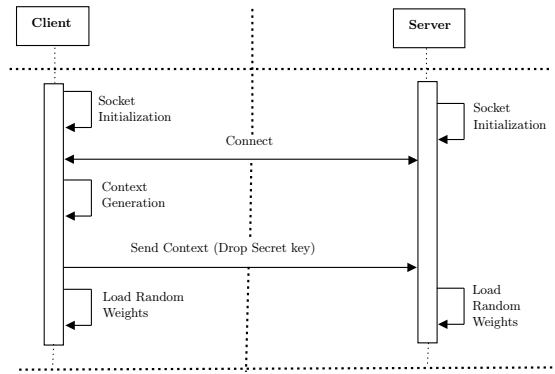
\begin{figure}
\centering
\begin{adjustbox}{width=0.6\textwidth, totalheight=0.4\textwidth}
\tikzset{every picture/.style={line width=0.75pt}}    

\begin{tikzpicture}[x=0.75pt,y=0.75pt,yscale=-1,xscale=1]

\draw   (113,266) -- (181,266) -- (181,301.5) -- (113,301.5) -- cycle ;

\draw  [dash pattern={on 0.84pt off 2.51pt}]  (149,303.5) -- (149,335.5) ;

\draw   (467,268) -- (534.06,268) -- (534.06,303.5) -- (467,303.5) -- cycle ;

\draw   (139,336.5) -- (157,336.5) -- (157,575.5) -- (139,575.5) -- cycle ;
\draw    (157.73,343.5) -- (183.35,343.5) ;
\draw    (183.35,343.5) -- (183.35,376.5) ;
\draw    (183.35,376.5) -- (161.85,376.5) ;
\draw [shift={(158.85,376.5)}, rotate = 360] [fill={rgb, 255:red, 0; green, 0; blue, 0 }  ][line width=0.08]  [draw opacity=0] (8.93,-4.29) -- (0,0) -- (8.93,4.29) -- cycle    ;
\draw    (157.7,424.5) -- (183.32,424.5) ;
\draw    (183.32,424.5) -- (183.32,457.5) ;
\draw    (183.32,457.5) -- (161.82,457.5) ;
\draw [shift={(158.82,457.5)}, rotate = 360] [fill={rgb, 255:red, 0; green, 0; blue, 0 }  ][line width=0.08]  [draw opacity=0] (8.93,-4.29) -- (0,0) -- (8.93,4.29) -- cycle    ; 
\draw    (157,504.49) -- (486,504.49) ;
\draw [shift={(489,504.5)}, rotate = 180.17] [fill={rgb, 255:red, 0; green, 0; blue, 0 }  ][line width=0.08]  [draw opacity=0] (8.93,-4.29) -- (0,0) -- (8.93,4.29) -- cycle    ;
\draw    (161,405.51) -- (487,405.51) ;
\draw [shift={(490,406.5)}, rotate = 180.17] [fill={rgb, 255:red, 0; green, 0; blue, 0 }  ][line width=0.08]  [draw opacity=0] (8.93,-4.29) -- (0,0) -- (8.93,4.29) -- cycle    ;
\draw [shift={(158,405.5)}, rotate = 0.17] [fill={rgb, 255:red, 0; green, 0; blue, 0 }  ][line width=0.08]  [draw opacity=0] (8.93,-4.29) -- (0,0) -- (8.93,4.29) -- cycle    ;
\draw    (157.7,524.5) -- (183.32,524.5) ;
\draw    (183.32,524.5) -- (183.32,557.5) ;
\draw    (183.32,557.5) -- (161.82,557.5) ;
\draw [shift={(158.82,557.5)}, rotate = 360] [fill={rgb, 255:red, 0; green, 0; blue, 0 }  ][line width=0.08]  [draw opacity=0] (8.93,-4.29) -- (0,0) -- (8.93,4.29) -- cycle    ;
\draw [color={rgb, 255:red, 0; green, 0; blue, 0 }  ,draw opacity=1 ][line width=1.5]  [dash pattern={on 1.69pt off 2.76pt}]  (115.5,598) -- (617.5,596) ;
\draw    (507.73,354.5) -- (533.35,354.5) ;
\draw    (533.35,354.5) -- (533.35,387.5) ;
\draw    (533.35,387.5) -- (511.85,387.5) ;
\draw [shift={(508.85,387.5)}, rotate = 360] [fill={rgb, 255:red, 0; green, 0; blue, 0 }  ][line width=0.08]  [draw opacity=0] (8.93,-4.29) -- (0,0) -- (8.93,4.29) -- cycle    ;
\draw  [dash pattern={on 0.84pt off 2.51pt}]  (498,304) -- (498,335) ;
\draw    (507.7,523.5) -- (533.32,523.5) ;
\draw    (533.32,523.5) -- (533.32,556.5) ;
\draw    (533.32,556.5) -- (511.82,556.5) ;
\draw [shift={(508.82,556.5)}, rotate = 360] [fill={rgb, 255:red, 0; green, 0; blue, 0 }  ][line width=0.08]  [draw opacity=0] (8.93,-4.29) -- (0,0) -- (8.93,4.29) -- cycle    ;
\draw [color={rgb, 255:red, 0; green, 0; blue, 0 }  ,draw opacity=1 ][line width=1.5]  [dash pattern={on 1.69pt off 2.76pt}]  (106.5,323) -- (614.5,323) ;
\draw  [dash pattern={on 0.84pt off 2.51pt}]  (147.9,574.5) -- (147,593.5) ;
\draw   (490,334.5) -- (508,334.5) -- (508,573.5) -- (490,573.5) -- cycle ;
\draw  [dash pattern={on 0.84pt off 2.51pt}]  (499.9,573.5) -- (499,592.5) ;
\draw [line width=1.5]  [dash pattern={on 1.69pt off 2.76pt}]  (334.5,268) -- (334.5,385) ;
\draw [line width=1.5]  [dash pattern={on 1.69pt off 2.76pt}]  (331,504) -- (328.5,600) ;
\draw [line width=1.5]  [dash pattern={on 1.69pt off 2.76pt}]  (333.5,409) -- (332.5,477) ;

\draw (540.61,353) node [anchor=north west][inner sep=0.75pt]   [align=left] {Socket \\Initialization};

\draw (539.61,519) node [anchor=north west][inner sep=0.75pt]   [align=left] {Load \\Random\\Weights};

\draw (188.35,346.5) node [anchor=north west][inner sep=0.75pt]   [align=left] {Socket \\Initialization};

\draw (308,383) node [anchor=north west][inner sep=0.75pt]   [align=left] {Connect};

\draw (192.61,525) node [anchor=north west][inner sep=0.75pt]   [align=left] {Load Random\\Weights};

\draw (191.61,425) node [anchor=north west][inner sep=0.75pt]   [align=left] {Context \\Generation};

\draw (219,480) node [anchor=north west][inner sep=0.75pt]   [align=left] {Send Context (Drop Secret key)};

\draw (479.97,277) node [anchor=north west][inner sep=0.75pt]   [align=left] {\textbf{Server}};

\draw (125.67,275) node [anchor=north west][inner sep=0.75pt]   [align=left] {\textbf{Client}};

\end{tikzpicture}

\end{adjustbox}

    \caption{Socket Initialization}
    \label{fig:socket}
\end{figure}

In the initialization phase, the client first connects to the server through the socket and 
receives 
a number of training parameters to synchronize (see~\autoref{fig:socket}). The parameters that are synchronized  between the client and server are  $E, \eta, n, N$. $E$ is the number of training epochs, $n$ is the batch size, $N$ is the number of batches to be trained and, $\eta$ represents the learning rate. These parameters should be synchronized on both sides in order to train the 1D CNN model in the same way (see \autoref{alg:client}, \autoref{alg:server}). In addition, both the client and server use the random weight initializer $\phi$ to initialize the weights $w_{i}$ and biases $b_{i}$ of their respective layers. The variables $\mathbf{a}^{(i)}$ (activation map), $\mathbf{z}^{(i)}$ (output of a tensor vector), and the gradients are also set to zero at the start of the phase. During this phase, the context is also generated, which holds the public key $\mathsf{pk}$ and secret key $\mathsf{sk}$ of the HE scheme as well as other parameters like polynomial modulus $\mathcal{P}$, coefficient modulus $\mathcal{C}$ and scaling factor $\Delta$.

\subsubsection{Forward propagation}
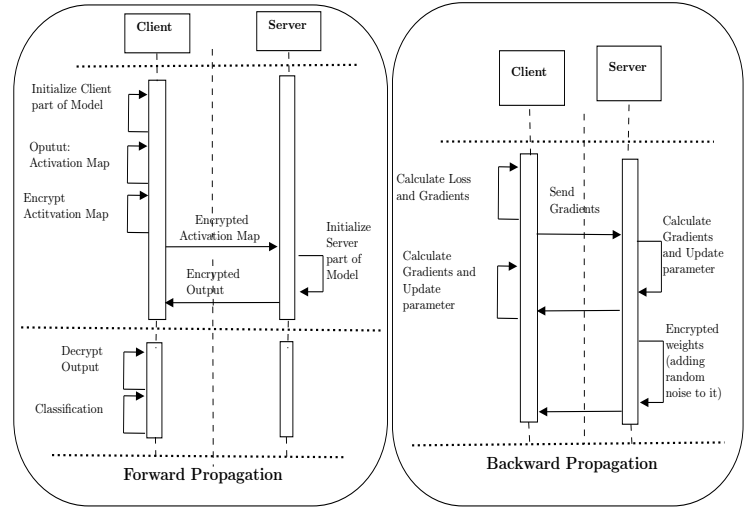
\begin{figure}
\centering
\begin{adjustbox}{width=0.8\textwidth, totalheight=0.55\textwidth}

\tikzset{every picture/.style={line width=0.75pt}} 

\begin{tikzpicture}[x=0.75pt,y=0.75pt,yscale=-1,xscale=1]

\draw [color={rgb, 255:red, 0; green, 0; blue, 0 }  ,draw opacity=1 ][line width=1.5]  [dash pattern={on 1.69pt off 2.76pt}]  (135.5,551) -- (430.8,549.48) ;
\draw    (210.15,610.6) -- (189.67,610.37) ;
\draw    (189.67,610.37) -- (190.05,576.42) ;
\draw    (190.05,576.42) -- (206.64,576.6) ;
\draw [shift={(209.64,576.63)}, rotate = 180.64] [fill={rgb, 255:red, 0; green, 0; blue, 0 }  ][line width=0.08]  [draw opacity=0] (8.93,-4.29) -- (0,0) -- (8.93,4.29) -- cycle    ;

\draw [color={rgb, 255:red, 0; green, 0; blue, 0 }  ,draw opacity=1 ][line width=1.5]  [dash pattern={on 1.69pt off 2.76pt}]  (77.5,789) -- (456.8,791.82) ;
\draw   (211,563.5) -- (228.8,563.5) -- (228.8,781.78) -- (211,781.78) -- cycle ;
\draw    (228.8,715.48) -- (342.8,715.29) ;
\draw [shift={(345.8,715.28)}, rotate = 179.9] [fill={rgb, 255:red, 0; green, 0; blue, 0 }  ][line width=0.08]  [draw opacity=0] (8.93,-4.29) -- (0,0) -- (8.93,4.29) -- cycle    ;
\draw   (348,559.5) -- (363.8,559.5) -- (363.8,781.78) -- (348,781.78) -- cycle ;
\draw    (367.73,723.5) -- (393.35,723.5) ;
\draw    (393.35,723.5) -- (393.35,756.5) ;
\draw    (393.35,756.5) -- (371.85,756.5) ;
\draw [shift={(368.85,756.5)}, rotate = 360] [fill={rgb, 255:red, 0; green, 0; blue, 0 }  ][line width=0.08]  [draw opacity=0] (8.93,-4.29) -- (0,0) -- (8.93,4.29) -- cycle    ;

\draw    (230.8,766.48) -- (347.8,766.28) ;
\draw [shift={(227.8,766.48)}, rotate = 359.9] [fill={rgb, 255:red, 0; green, 0; blue, 0 }  ][line width=0.08]  [draw opacity=0] (8.93,-4.29) -- (0,0) -- (8.93,4.29) -- cycle    ;
\draw  [dash pattern={on 4.5pt off 4.5pt}]  (277.5,535) -- (277.5,707) ;
\draw  [dash pattern={on 4.5pt off 4.5pt}]  (278,727) -- (278.5,771) ;
\draw  [dash pattern={on 4.5pt off 4.5pt}]  (278,770) -- (278.5,916) ;
\draw  [dash pattern={on 4.5pt off 4.5pt}]  (219.5,539) -- (219.8,564.33) ;
\draw  [dash pattern={on 4.5pt off 4.5pt}]  (357.5,538) -- (357.8,560.33) ;
\draw  [dash pattern={on 4.5pt off 4.5pt}]  (219,783) -- (218.77,794.48) -- (218.5,808) ;
\draw  [dash pattern={on 4.5pt off 4.5pt}]  (357,783) -- (356.5,808) ;
\draw   (186,503) -- (254,503) -- (254,538.5) -- (186,538.5) -- cycle ;

\draw   (324,501) -- (391.06,501) -- (391.06,536.5) -- (324,536.5) -- cycle ;

\draw   (209,803) -- (224.8,803) -- (224.8,889.78) -- (209,889.78) -- cycle ;
\draw   (348.5,802) -- (361.8,802) -- (361.8,888.78) -- (348.5,888.78) -- cycle ;
\draw [color={rgb, 255:red, 0; green, 0; blue, 0 }  ,draw opacity=1 ][line width=1.5]  [dash pattern={on 1.69pt off 2.76pt}]  (109.8,905.48) -- (418.15,907.41) ;
\draw  [dash pattern={on 4.5pt off 4.5pt}]  (218,889) -- (218.5,910) ;
\draw  [dash pattern={on 4.5pt off 4.5pt}]  (355,889) -- (355.5,910) ;
\draw    (616.8,703.78) -- (702.3,704.41) ;
\draw [shift={(705.3,704.44)}, rotate = 180.42] [fill={rgb, 255:red, 0; green, 0; blue, 0 }  ][line width=0.08]  [draw opacity=0] (8.93,-4.29) -- (0,0) -- (8.93,4.29) -- cycle    ;
\draw    (721.73,710.43) -- (747.35,710.43) ;
\draw    (747.35,710.43) -- (747.35,756.75) ;
\draw    (747.35,756.75) -- (725.85,756.75) ;
\draw [shift={(722.85,756.75)}, rotate = 360] [fill={rgb, 255:red, 0; green, 0; blue, 0 }  ][line width=0.08]  [draw opacity=0] (8.93,-4.29) -- (0,0) -- (8.93,4.29) -- cycle    ;

\draw [color={rgb, 255:red, 0; green, 0; blue, 0 }  ,draw opacity=1 ][line width=1.5]  [dash pattern={on 1.69pt off 2.76pt}]  (486.5,896.59) -- (813.3,893.95) ;
\draw   (599.8,630.69) -- (617.7,630.69) -- (617.7,875.33) -- (599.8,875.33) -- cycle ;
\draw    (620,773.96) -- (703.8,773.48) ;
\draw [shift={(617,773.98)}, rotate = 359.67] [fill={rgb, 255:red, 0; green, 0; blue, 0 }  ][line width=0.08]  [draw opacity=0] (8.93,-4.29) -- (0,0) -- (8.93,4.29) -- cycle    ;
\draw [color={rgb, 255:red, 0; green, 0; blue, 0 }  ,draw opacity=1 ][line width=1.5]  [dash pattern={on 1.69pt off 2.76pt}]  (516.5,619.41) -- (802.3,619.57) ;
\draw  [dash pattern={on 4.5pt off 4.5pt}]  (666.5,594.14) -- (666.8,895.48) ;
\draw  [dash pattern={on 4.5pt off 4.5pt}]  (609.5,589.93) -- (610.5,633.44) ;
\draw  [dash pattern={on 4.5pt off 4.5pt}]  (713.5,585.72) -- (712.5,629.23) ;
\draw   (578,536.6) -- (646,536.6) -- (646,586.42) -- (578,586.42) -- cycle ;

\draw   (680,532.39) -- (747.06,532.39) -- (747.06,582.21) -- (680,582.21) -- cycle ;

\draw   (706.5,635.55) -- (722.8,635.55) -- (722.8,877.33) -- (706.5,877.33) -- cycle ;
\draw  [dash pattern={on 4.5pt off 4.5pt}]  (608.5,875.3) -- (609.8,898.33) ;
\draw    (211.15,702.6) -- (188.63,702.35) ;
\draw    (188.63,702.35) -- (189.01,668.39) ;
\draw    (189.01,668.39) -- (207.55,668.6) ;
\draw [shift={(210.55,668.63)}, rotate = 180.64] [fill={rgb, 255:red, 0; green, 0; blue, 0 }  ][line width=0.08]  [draw opacity=0] (8.93,-4.29) -- (0,0) -- (8.93,4.29) -- cycle    ;

\draw    (210.15,657.6) -- (188.67,657.36) ;
\draw    (188.67,657.36) -- (189.05,623.4) ;
\draw    (189.05,623.4) -- (206.6,623.6) ;
\draw [shift={(209.6,623.63)}, rotate = 180.64] [fill={rgb, 255:red, 0; green, 0; blue, 0 }  ][line width=0.08]  [draw opacity=0] (8.93,-4.29) -- (0,0) -- (8.93,4.29) -- cycle    ;

\draw    (207.15,845.6) -- (184.63,845.35) ;
\draw    (184.63,845.35) -- (185.01,811.39) ;
\draw    (185.01,811.39) -- (203.55,811.6) ;
\draw [shift={(206.55,811.63)}, rotate = 180.64] [fill={rgb, 255:red, 0; green, 0; blue, 0 }  ][line width=0.08]  [draw opacity=0] (8.93,-4.29) -- (0,0) -- (8.93,4.29) -- cycle    ;

\draw    (207.15,885.6) -- (184.63,885.35) ;
\draw    (184.63,885.35) -- (185.01,851.39) ;
\draw    (185.01,851.39) -- (203.55,851.6) ;
\draw [shift={(206.55,851.63)}, rotate = 180.64] [fill={rgb, 255:red, 0; green, 0; blue, 0 }  ][line width=0.08]  [draw opacity=0] (8.93,-4.29) -- (0,0) -- (8.93,4.29) -- cycle    ;

\draw   (69.8,573.03) .. controls (69.8,529.84) and (104.81,494.83) .. (148,494.83) -- (382.6,494.83) .. controls (425.79,494.83) and (460.8,529.84) .. (460.8,573.03) -- (460.8,873.67) .. controls (460.8,916.86) and (425.79,951.87) .. (382.6,951.87) -- (148,951.87) .. controls (104.81,951.87) and (69.8,916.86) .. (69.8,873.67) -- cycle ;
\draw   (466,566.14) .. controls (466,525.63) and (498.84,492.78) .. (539.36,492.78) -- (759.44,492.78) .. controls (799.96,492.78) and (832.8,525.63) .. (832.8,566.14) -- (832.8,878.51) .. controls (832.8,919.02) and (799.96,951.87) .. (759.44,951.87) -- (539.36,951.87) .. controls (498.84,951.87) and (466,919.02) .. (466,878.51) -- cycle ;
\draw    (598.15,690.44) -- (576.67,690.11) ;
\draw    (576.67,690.11) -- (577.05,642.45) ;
\draw    (577.05,642.45) -- (594.6,642.73) ;
\draw [shift={(597.6,642.77)}, rotate = 180.9] [fill={rgb, 255:red, 0; green, 0; blue, 0 }  ][line width=0.08]  [draw opacity=0] (8.93,-4.29) -- (0,0) -- (8.93,4.29) -- cycle    ;

\draw    (597.15,780.92) -- (575.67,780.59) ;
\draw    (575.67,780.59) -- (576.05,732.93) ;
\draw    (576.05,732.93) -- (593.6,733.2) ;
\draw [shift={(596.6,733.25)}, rotate = 180.9] [fill={rgb, 255:red, 0; green, 0; blue, 0 }  ][line width=0.08]  [draw opacity=0] (8.93,-4.29) -- (0,0) -- (8.93,4.29) -- cycle    ;

\draw  [dash pattern={on 4.5pt off 4.5pt}]  (715.5,877.3) -- (716.8,900.33) ;
\draw    (622,865.96) -- (705.8,865.48) ;
\draw [shift={(619,865.98)}, rotate = 359.67] [fill={rgb, 255:red, 0; green, 0; blue, 0 }  ][line width=0.08]  [draw opacity=0] (8.93,-4.29) -- (0,0) -- (8.93,4.29) -- cycle    ;
\draw    (723.73,801.48) -- (749.35,801.48) ;
\draw    (749.35,801.48) -- (749.35,857.48) ;
\draw    (749.35,857.48) -- (727.85,857.48) ;
\draw [shift={(724.85,857.48)}, rotate = 360] [fill={rgb, 255:red, 0; green, 0; blue, 0 }  ][line width=0.08]  [draw opacity=0] (8.93,-4.29) -- (0,0) -- (8.93,4.29) -- cycle    ;

\draw (87.35,565.5) node [anchor=north west][inner sep=0.75pt]   [align=left] {Initialize Client \\part of Model};
\draw (84.61,618) node [anchor=north west][inner sep=0.75pt]   [align=left] {Oputut:\\Activation Map};
\draw (78.61,665) node [anchor=north west][inner sep=0.75pt]   [align=left] {Encrypt\\Actitvation Map};
\draw (241.73, 685) node [anchor=north west][inner sep=0.75pt]   [align=left] { \ \ \ \ Encrypted \\Activation Map};
\draw (395.35,691.5) node [anchor=north west][inner sep=0.75pt]   [align=left] {Initialize \\Server \\part of\\ Model};
\draw (249,734) node [anchor=north west][inner sep=0.75pt]   [align=left] {Encrypted \\Output};
\draw (118.35,804) node [anchor=north west][inner sep=0.75pt]   [align=left] {Decrypt \\Output};
\draw (90.61,857.5) node [anchor=north west][inner sep=0.75pt]   [align=left] {Classification};
\draw (468.61,647.95) node [anchor=north west][inner sep=0.75pt]   [align=left] {Calculate Loss \\and Gradients};
\draw (629,659.76) node [anchor=north west][inner sep=0.75pt]   [align=left] {Send \\Gradients};
\draw (747.35,685.5) node [anchor=north west][inner sep=0.75pt]   [align=left] {Calculate\\Gradients \\and Update \\parameter};
\draw (475.35,718.13) node [anchor=north west][inner sep=0.75pt]   [align=left] {Calculate\\Gradients and \\Update \\parameter};
\draw (183,915.22) node [anchor=north west][inner sep=0.75pt]   [align=left] {\textbf{{\large Forward Propagation}}};
\draw (563,905.68) node [anchor=north west][inner sep=0.75pt]   [align=left] {\textbf{{\large Backward Propagation}}};
\draw (750.35,784.5) node [anchor=north west][inner sep=0.75pt]   [align=left] {Encrypted \\weights \\(adding \\random \\noise to it)};
\draw (690.97,545.45) node [anchor=north west][inner sep=0.75pt]   [align=left] {\textbf{Server}};
\draw (588.67,549.66) node [anchor=north west][inner sep=0.75pt]   [align=left] {\textbf{Client}};
\draw (334.97,507) node [anchor=north west][inner sep=0.75pt]   [align=left] {\textbf{Server}};
\draw (196.67,509) node [anchor=north west][inner sep=0.75pt]   [align=left] {\textbf{Client}};

\end{tikzpicture}

\end{adjustbox}

    \caption{Forward and Backward Propagation}
    \label{fig:forward}
\end{figure}

The forward propagation starts from the client side. As can be seen in
~\autoref{fig:forward}, the client and server initialize their part of the model. The client first generates the batch $(x, y)$, which has $n$ data examples extracted from the train set $D$. $x$ represents the input data, and $y$ shows their labels. During the forward propagation phase, the client forward propagates the input $x$ (see~\autoref{equ: forward}) until 
layer $l$ and generates an activation map $(a^{i})$. 
\begin{equation}
\begin{aligned}
\label{equ: forward}
    z^{i} \leftarrow f^{( i)} ( \mathbf{a}^{( i-1)}) \\
    a^{i}  \leftarrow  g^{( i)} ( \mathbf{z}^{( i)})
\end{aligned}
\end{equation}
where $f^{i}$ and $g^{i}$ are the convolution operation of layer \textit{i} and activation function of layer \textit{i} respectively.

The client encrypts the activation map and sends the encrypted activation map ($\overbar{\mathbf{a}^{(l)}}$) to the server (see~\autoref{alg:client}). The server performs computation on the encrypted activation map and sends the output to the client ($\overbar{\mathbf{a}^{(L)}}$). Finally, the client executes the final softmax layer of the 1D CNN model and calculates the loss function ($J$) between the original $y$ and the predicted output $\hat{y}$. The loss function produces error $J$ by computing the cross entropy loss.

\subsubsection{Backward propagation}

As can be seen in~\autoref{fig:forward}, the backpropagation starts from the client side. The client calculates the loss function, the gradients of the loss function w.r.t the activation map and send these gradients to the server. On receiving the gradients from the client, the server calculates the gradients of the loss function w.r.t its linear layer and update the parameter of this layer. More specifically, as shown in \autoref{alg:client} and \autoref{alg:server}, the client calculates and sends the gradients of error w.r.t $a^{(L)}$ to the server. The server continues the backpropagation, calculates $\frac{dJ}{db}$, $\overline{\frac{dJ}{dw'}}$ and $\frac{dJ}{da}$. The server then sends $\frac{dJ}{da}$ to the client. After receiving the gradients, the backpropagation continues to the first hidden layer on the client side. Note that as $\frac{dJ}{dw'}$ becomes encrypted, the server's weights will become encrypted after updating the parameters as in~\autoref{equ:backward}. After updating this parameters for several forward and backward pass, the noise level in $w$ can cause overflow, rendering the computation in correct. To solve this issue, the server also sends the encrypted weights to the client after adding some fixed noise to it to preserve the weight's privacy. Upon reception, the client decrypts the weights to reset the HE noise level and sends the weights back to the server. After receiving the decrypted weights, the server subtract the fixed noise added earlier and continues to the next forward propagation pass.
\begin{equation}
\begin{aligned}
\label{equ:backward}
    \overbar{\left( \boldsymbol{w}^{{'}(L)} \right)}= HE.Enc(\boldsymbol{w}^{{'}(L)}) \\
    \overbar{\left( \boldsymbol{w}^{{'}(L)} \right)}= \overbar{\left( \boldsymbol{w}^{{'}(L)} \right)}-\overbar{\frac{\partial J}{\partial \boldsymbol{w}^{{'}(L)}}}
\end{aligned}
\end{equation}
The difference between our new proposed protocol and the protocol in algorithms~3 and 4 of~\cite{khan2023split} is that instead of sending both $\frac{\partial J}{\partial \mathbf{a}^{(L)}}$ and $\frac{\partial J}{\partial \boldsymbol{w}^{(L)}}$ to the server, in our protocol's backward pass, the client only sends $\frac{\partial J}{\partial \mathbf{a}^{(L)}}$ to the server, hence resolving the privacy leakage reported in~\autoref{sec:privacyLeakageAnal}.

\section{Performance Analysis}
\label{sec:performance}
In this section, we describe the experimental setups, datasets, and results from training our framework.

\subsection{Evaluation} 
\label{subsec:evaluation}

\noindent \textbf{Experimental Setup:} 
To process the data and train the neural networks, we use a machine with Intel Core i7-8700 CPU processor, 32 Gb of RAM, GeForce GTX 1070 Ti GPU with 8 Gb of GPU memory and running Ubuntu 20.04 LTS. The algorithms are implemented using Python 3.9.7, 
the PyTorch framework version 1.8.1+cu102 for constructing the neural network, 
 the TenSeal framework~\cite{tenseal2021} version 0.3.10 for HE and are evaluated on the local network. 

\begin{figure}
	\centering
	\includegraphics[width=0.7\textwidth, totalheight=0.25\textwidth]{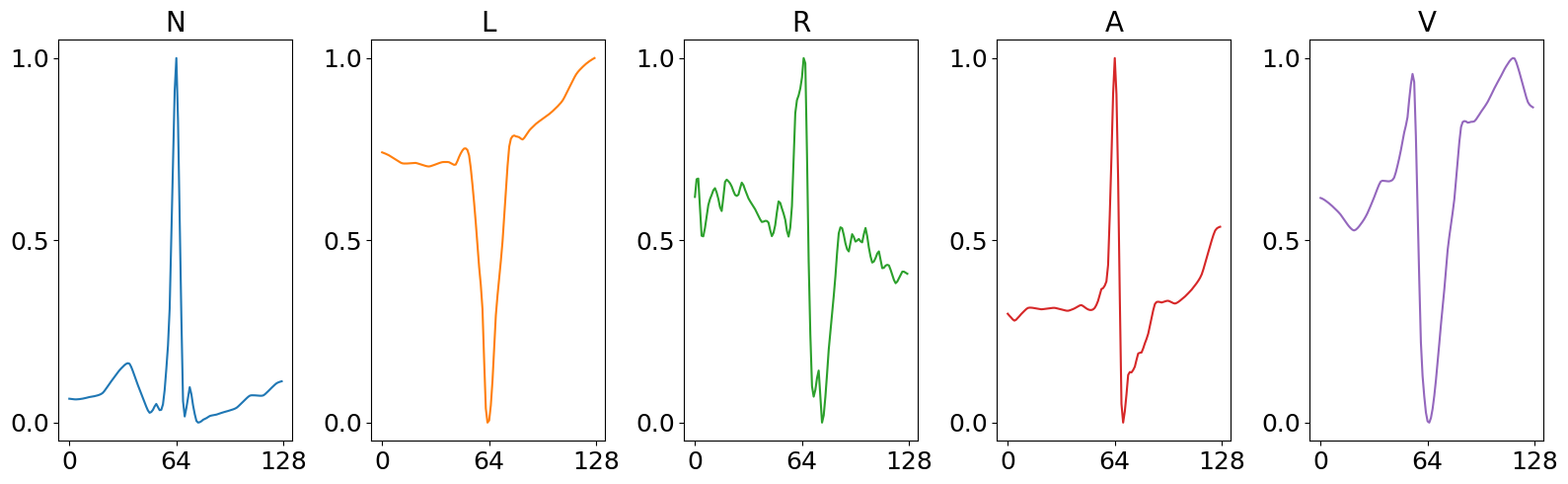}
	\caption{Example heartbeats from the MIT-BIH dataset.}
	\label{fig:mitbih_signals}
\end{figure}

\noindent \textbf{Datasets}
We evaluate the performance of our framework on two Electrocardiography (ECG) datasets: the MIT-BIH dataset~\cite{moody2001impact} and the PTB-XL dataset~\cite{wagner2020ptb}.
\paragraph{MIT-BIH} This dataset contains 48 half-hour excerpts of two-channel ECG recordings, obtained from 47 subjects from 1975 to 1979. 
In our work, we use the processed version of the dataset from~\cite{abuadbba2020can}, which contains 26,490 heartbeat samples. Each sample belongs to one out of five classes: Normal (N), Left bundle branch block (L), Right bundle branch block (R), Atrial premature contraction (A), Ventricular premature contraction (V). 
~\autoref{fig:mitbih_signals} shows some example signals from the processed MIT-BIH datset. To train the NNs, the dataset is split into  a 50-50 ratio. A train or test split contains 13,245 examples, each example is a time series signal of length 128.

\begin{figure}
	\centering
	\includegraphics[width=0.7\textwidth, totalheight=0.4\textwidth]{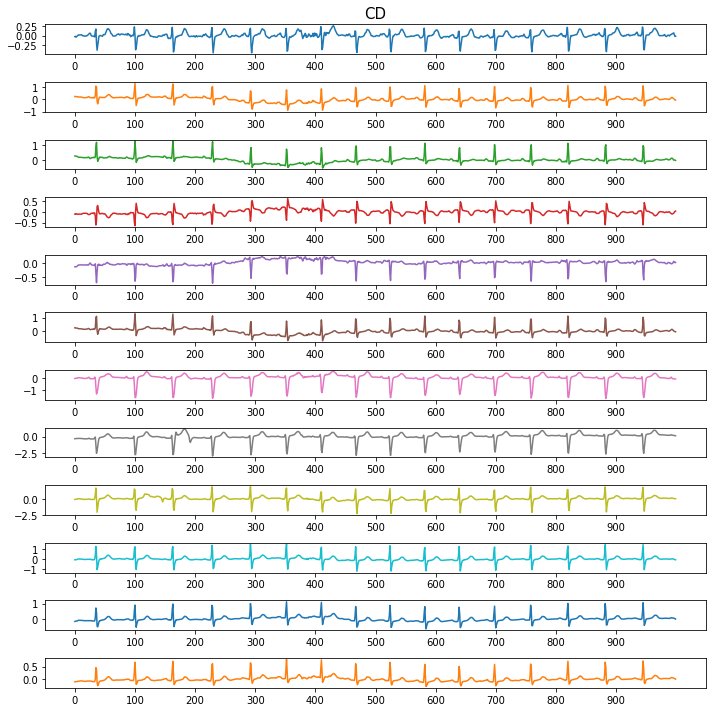}
	\caption{An example CD ECG signal from the PTB-XL dataset.}
	\label{fig:ptbxl_cd_signal}
\end{figure}

\paragraph{PTB-XL} It is a large ECG dataset that contains 21837 clinical ECGs from 18885 patients, collected between 1989 and 1996. Each ECG waveform has 12 channels (leads) and is 10 seconds long. The raw waveforms are sampled at two different sampling rates (100 Hz and 500 Hz) and 
annotated by up to two cardiologists, who may assign multiple statements to one ECG signal. An ECG signal can be diagnosed to belong to one out of five superclasses: Normal ECG (NORM), Myocardial Infarction (MI), ST/T Change (STTC), Conduction Disturbance (CD), and Hypertrophy (HYP).~\autoref{fig:ptbxl_cd_signal} shows an example of an ECG signal that belongs to the CD class; we can see that there are a total of 12 time series signals in the figure, as 
we are dealing with 12-lead ECG signals. The dataset is processed according to~\cite{khan2023split}, where 
a sampling rate of 100 Hz is used. Following the example of ~\cite{khan2023split}, if an ECG signal belongs to more than one classes, we only choose one 
(the first class). The whole dataset is then split into train-test splits 
at a ratio of 90\%-10\%. More specifically, the training split contains 19,267 data samples and the test split contains 2,163 data samples. 

\begin{minipage}{0.5\textwidth}
\begin{algorithm}[H]
\scriptsize
\SetAlgoLined
 \textbf{Initialization:}\\
 $s\leftarrow$ socket initialization\;
 \textit{s.connect}\\
 $\eta, n, N, E \leftarrow s.synchronize()$\\
 $ \{\boldsymbol{w}^{( i)}, \boldsymbol{b}^{( i)}\}_{\forall i\in \{0..l\}} \ \leftarrow  \Phi $\\
 $\{\mathbf{z}^{( i)}\}_{\forall i\in \{0..l\}} ,\{\mathbf{a}^{( i)}\}_{\forall i\in \{0..l\}}\leftarrow \emptyset \ $\\
 $ \left\{\frac{\partial J}{\partial \mathbf{z}^{i}}\right\}_{ i\in \{0..l\}} ,\left\{\frac{\partial J}{\partial \mathbf{a}^{i}}\right\}_{i\in \{0..l\}}\leftarrow \emptyset \ $\\
  \textbf{Context Initialization:}\\
 $\displaystyle  \mathsf{ctx_{pri}},\ \leftarrow \ \mathcal{P}, \ \mathcal{C}, \ \Delta, \  \mathsf{pk}, \ \mathsf{sk}$\\
 $\displaystyle  \mathsf{ctx_{pub}},\ \leftarrow \ \mathcal{P}, \ \mathcal{C}, \ \Delta, \  \mathsf{pk}$\\
			 $\displaystyle  s.send( \mathsf{ctx_{pub}})$\\
	\For{$\displaystyle e \ \text{in} \ E $}{
	\For{$\displaystyle \text{ each} \ \text{batch}\ ( \mathbf{x},\ \mathbf{y}) \ \text{ from}\ \mathbf{D}\ $}{
	$\displaystyle  \mathbf{Forward\ propagation:}$\\
	$\displaystyle  O.zero\_grad()  $\\
	$\displaystyle  \mathbf{a}^{0}\ \ \leftarrow \mathbf{x}$ \\
	\For{$i\ \leftarrow \ 1\ \mathbf{to} \ l$}{
	$\displaystyle \ \ \ \ \mathbf{z}^{( i)} \ \leftarrow \ f^{( i)}\left( \mathbf{a}^{( i-1)}\right)$\\
	$\displaystyle \ \ \ \ \mathbf{a}^{i} \ \leftarrow \ g^{( i)}\left( \mathbf{z}^{( i)}\right)$}
	$\displaystyle \overbar{\mathbf{a}^{(l)}} \ \leftarrow \ \mathsf{HE.Enc}\left(\mathsf{pk}, \mathbf{a}^{(l)}\right)$\\
	$\displaystyle s.send \ \overbar{(\mathbf{a}^{(l)})}$\\
	$\displaystyle  s.receive\ ( \overbar{\mathbf{a}^{(L)})}$\\
	$\displaystyle  \mathbf{a}^{( L)} \ \leftarrow \ \mathsf{HE.Dec}\left(\mathsf{sk}, \overbar{\mathbf{a}^{( L)}}\right)$\\
	$\displaystyle  \hat{\mathbf{y}} \ \leftarrow \ Softmax\left(\mathbf{a}^{( L)}\right)$\\
	$\displaystyle  \mathbf{J} \leftarrow \mathcal{L} (\hat{\mathbf{y}}, \mathbf{y})$\\
	$\displaystyle \mathbf{Backward\ propagation:}$\\
	$\displaystyle \text{Compute}\left\{\frac{\partial J}{\partial \hat{\mathbf{y}}} \& \frac{\partial J}{\partial \mathbf{a}^{(L)} } \right\}$\\
	$\displaystyle s.send\left(\frac{\partial J}{\partial \mathbf{a}^{(L)} } \right)$\\
	$\displaystyle s.receive\left( \frac{\partial J}{\partial \mathbf{a}^{(l)}} \right)$\\
	\For{$i\leftarrow l\ \text{down to} \ 1$}{
	$\displaystyle  \ \ \ \ \text{Compute} \left\{ \frac{\partial J}{\partial \boldsymbol{w}^{(i)}}, \ \frac{\partial J}{\partial \boldsymbol{b}^{(i)}} \right\}$\\
	$\displaystyle \ \ \ \ \text{Update}\ \boldsymbol{w}^{( i)},\ \boldsymbol{b}^{(i)} $
 }
 $\displaystyle \boldsymbol{w}^{{'}(L)} = HE.Dec\overbar{\left( \boldsymbol{w}^{{'}(L)} \right)} $\\
 $\displaystyle s.send \left( \boldsymbol{w}^{{'}(L)}\right) $
	}	
	}
 \caption{\textbf{Client Side}}
 \label{alg:client}
\end{algorithm}	
\end{minipage}
\begin{minipage}{0.5\textwidth}
\begin{algorithm}[H]
\scriptsize
\SetAlgoLined
 \textbf{Initialization:}\\
 $s\leftarrow$ socket initialization\;
 \textit{s.connect}\\
 $\eta, n, N, E \leftarrow s.synchronize()$\\
 $ \{\boldsymbol{w}^{( i)}, \boldsymbol{b}^{( i)}\}_{\forall i\in \{0..l\}} \ \leftarrow \Phi $\\
 $\displaystyle  \{\mathbf{z}^{( i)}\}_{\forall i\in \{l+1..L\}} \leftarrow \emptyset \ $\\
 $\displaystyle  \left\{\frac{\partial J}{\partial \mathbf{z}^{( i)}}\right\}_{\forall i\in \{l+1..L\}} \leftarrow \emptyset \ $\\
 \For{$\displaystyle e \ \in \ E $}{
 	\For{$\displaystyle i \leftarrow 1 \ \mathbf{to} \ N \ $}{
 	$\displaystyle \mathbf{Forward\ propagation:}$\\
 	$\displaystyle O.zero\_grad()  $\\
 	$\displaystyle s.receive\ (\overbar{(\mathbf{a}^{(l)})}) \ \ $ \\
$\displaystyle \overbar{\mathbf{a}^{(L)}} \ \leftarrow \ \left(\overbar{(\mathbf{a}^{(l)})} . \ w^{'} + b \right)$\\
$\displaystyle s.send\left( \overbar{\mathbf{a}^{(L)}}\right)$\\
$\displaystyle \mathbf{Backward\ propagation:}$\\
$\displaystyle s.receive \ \left( \frac{\partial J}{\partial \mathbf{a}^{(L)}}\right)$\\
$\displaystyle \text{Compute}\ \left\{ \overbar{\frac{\partial J}{\partial \boldsymbol{w}^{{'}(L)}}}, \ \frac{\partial J}{\partial \boldsymbol{b}^{(L)}} , \ \frac{\partial J}{\partial \boldsymbol{a}^{(L)}} \right\}$\\
$\displaystyle \text{Encrypt and add noise to}\ \boldsymbol{w}^{{'}(L)}$  \\
$\displaystyle \text{Update}~\boldsymbol{w}^{{'}(L)}, ,\ \boldsymbol{b}^{(L)}.$ \\
$\displaystyle \text{Compute}\ \frac{\partial J}{\partial \mathbf{a}^{( l)}} $\\
$\displaystyle s.send \left( \frac{\partial J}{\partial \mathbf{a}^{( l)}} \right)$\\
$\displaystyle s.send \overbar{\left( \boldsymbol{w}^{{'}(L)} \right)}$\\
$\displaystyle s.receive \left( \boldsymbol{w}^{{'}(L)}\right) $ \\ 
$\displaystyle \text{Subtract noise from}~\boldsymbol{w}^{{'}(L)} $
   }
 }
 \caption{\textbf{Server Side}}
 \label{alg:server}
\end{algorithm}
\end{minipage}
 
\noindent \textbf{Hyper- and HE- parameters settings:} 
First, we reproduced the 
training results for the local version and the split version on HESplit network plaintext data 
from~\cite{khan2023split} and 
concluded in the same results, 
namely 88.06\% accuracy on the MIT-BIH dataset. We also trained a plaintext split network on the PTB-XL dataset and got 67.68\% accuracy. Note that these accuracies are low, however, as we only focus on comparing the accuracies when training on plaintext versus encrypted data, getting high predictive accuracies on these datasets is not the focus of this work. To train the network on these datasets, we use 10 epochs with a learning rate of 0.001 and batch size 4. For plaintext training, we used the Adam optimizer~\cite{kingma2014Adam} on both the client and server side.

We use the same NN hyperparameters when training the HESplit network on encrypted activation maps, however, the optimizer on the server side is a simple stochastic gradient descent. To encrypt the activation maps before sending them to the server, we employ two sets of HE parameters as following:
\begin{enumerate}
    \item S1: $N=2^{13}=8192,  ~\mathcal{C}=[40, 21, 21, 21, 40], ~\Delta=2^{21}$
    \item S2: $N=2^{14}=16384, ~\mathcal{C}=[40, 21, 21, 21, 40], ~\Delta=2^{21}$
\end{enumerate}
The only difference between the two sets of HE parameters is the polynomial modulus degree $N$, which directly affects the computational performance of the scheme (bigger is worse) and its security level (bigger is better). Apart from that, the coefficient modulus $\mathcal{C}$ is a list of prime numbers that define scheme
noise levels and HE multiplication depth. Each HE multiplication consumes a prime in $\mathcal{C}$, and after all primes in the list are consumed, we will no longer be able to perform multiplication in the encrypted domain. $\Delta$ is called the scaling factor, which is the factor where the plaintext message is multiplied by to maintain a certain level of precision during the encoding process. We vary only $N$ to assess the trade-off between security and performance of our protocol.
The results 
of training the HESplit network with S1 and S2 on the MIT-BIH dataset can be found in~\autoref{tab:hesplitResults}. 
Due to computational limitations, we can only train HESplit with $\mathcal{S}_1$ on the PTB-XL dataset.

\begin{table}
\centering
\caption{Training results of HESplit network on the MIT-BIH abd PTB-XL dataset using different sets of HE parameters and batch sizes. The training duration (in sec) and communication (in Gb) overhead are reported as the average value per epoch}
\label{tab:hesplitResults}
\scalebox{0.74}{
\begin{tblr}{
  row{3} = {c},
  row{4} = {c},
  row{5} = {c},
  row{6} = {c},
  row{7} = {c},
  row{8} = {c},
  row{9} = {c},
  row{10} = {c},
  row{11} = {c},
  row{13} = {c},
  row{14} = {c},
  row{15} = {c},
  row{16} = {c},
  cell{1}{1} = {frenchblue,c,fg=white},
  cell{1}{2} = {frenchblue,c,fg=white},
  cell{1}{3} = {frenchblue,c,fg=white},
  cell{1}{4} = {frenchblue,c,fg=white},
  cell{1}{5} = {frenchblue,c,fg=white},
  cell{1}{6} = {frenchblue,c,fg=white},
  cell{2}{1} = {r=10}{},
  cell{2}{2} = {r=5}{c},
  cell{2}{3} = {c},
  cell{2}{4} = {c},
  cell{2}{5} = {c},
  cell{2}{6} = {c},
  cell{7}{2} = {r=5}{},
  cell{12}{1} = {r=5}{},
  cell{12}{2} = {r=5}{c},
  cell{12}{3} = {c},
  cell{12}{4} = {c},
  cell{12}{5} = {c},
  cell{12}{6} = {c},
  vlines,
  hline{1-2,12,17} = {-}{},
  hline{3-6,8-11,13-16} = {3-6}{},
  hline{7} = {2-6}{},
}
        Dataset  &  HE param set  & Batch Size & Accuracy (\%) & Training Time (s) & Communication (Gb) \\
\textbf{MIT-BIH } & S1 & 4~         & 83.49~        & 8100.60           & 239.53             \\
                  &    & 8~         & 78.01         & 4780.73~         & 121.57             \\
                  &    & 16         & 74.77        & 2855.29          & 62.58             \\
                  &    & 32         & 73.79        & 1658.73          & 33.09             \\
                  &    & 64         & 62.08        & 987.42           & 18.35             \\
                  & S2 & 4          & 83.09         & 19746.85         & 471.57            \\
                  &    & 8          & 67.94         & 10378.92         & 239.33            \\
                  &    & 16         & 70.33         & 5358.59          & 123.20            \\
                  &    & 32         & 72.47         & 3152.82          & 65.14             \\
                  &    & 64         & 64.42         & 1913.11          & 36.11             \\
\textbf{PTB-XL}   & S1 & 4          & 58.71         & 100060.60        & ~2624.85           \\
                  &    & 8          & 58.95         & 49334.95~        & 1315.38            \\
                  &    & 16         & 57.10         & 22501.79         & 660.65            \\
                  &    & 32         & 59.36         & 12370.38         & 333.23            \\
                  &    & 64         & 56.45         & 5702.29          & 169.60            
\end{tblr}
}
\end{table}

For the MIT-BIH dataset, the best test accuracy, after training on encrypted activation maps, is 83.49\% when using S1 with the training batch size of 4. Overalls, we can see that for MIT-BIH, S1 achieve better accuracy than S2, and the training time as well as communication cost using S1 are also about half when using S2. This indicates the trade off between security and utility. For both S1 and S2, the batch size of 4 achieve the best accuracies overall, since once the batch size get bigger, more compression will be done in HE batch processing and more noise will be introduced in HE batch computations. Nonetheless, the bigger the batch size, the less training time and communication needed, and this relationship is linear as observed from our experiments. When the batch size is small ($n=4$), S1 and S2 produce very comparable accuracies (83.49\% vs 83.09\%), and S2 incur double training time as well as communication cost compared to S1. However, when $n$ gets bigger, the accuracies of S2 degrade very quickly compared to S1, as can observed when $n=16$ and $n=32$. When $n=64$, both S1 and S2 produce very low predictive power with 62.08\% and 64.42\%. We reason that when $n$ gets big, the accumulated noise also become very big that the neural network does equally bad on both cases. For the PTB-XL dataset, accuracies across all batch sizes vary much less compared to those from MIT-BIH. The reason for this is that because an ECG example from PTB-XL is much bigger than that from MIT-BIH (length 1000 vs length 128), compressing along the batch size when doing HE encryption and noise incurred during HE computation produce less effects on the results.

In~\autoref{tab:comparisons}, we provide a comparison between HESplit network, the work from~\cite{khan2023split} (Split Ways) and the split plaintext training results when the training batch size is 4. In terms of accuracy, for MIT-BIH, HESplit network produces 4.57\% and 1.82\% less than the plaintext version and the Split Ways network. For PTB-XL, the difference is bigger: 6.71\% less than Split Ways and 8.97\% less than plaintext. W.r.t training time and communication, both HESplit and Split Ways require thousands of time more to train compared to the plaintext version. However, compared to Split Ways, HESplit is 6$\times$ faster and about 160$\times$ less communication overhead on MIT-BIH. On PTB-XL, HESplit requires 1.3$\times$ more time to train but 44$\times$ less communication overhead compared to Split Ways. The reason for HESplit having longer training time compared to Split Ways on PTB-XL is that the batch size is only small ($n=4$). We predict when $n$ get bigger, we will achieve shorter training time compared to Split Ways. On the other hand, we can see that overalls, HESplit produces much less communication cost compared to Split Ways. 

\begin{table}
\centering
\caption{. The training time (in seconds) and communication (in Gb) overhead are reported as the average value per epoch.}
\label{tab:comparisons}
\scalebox{0.74}{
\begin{tblr}{
  row{1} = {frenchblue,fg=white},
  cell{2}{1} = {r=3}{},
  cell{5}{1} = {r=3}{},
  vlines,
  hline{1-2,5,8} = {-}{},
  hline{3-4,6-7} = {2-6}{},
}
Dataset          & Framework   & Batch Size & Accuracy (\%) & Training Time (s) & Communication (Gb) \\
\textbf{MIT-BIH} & HESplit  & 4~         & 83.49~        & 8100.60           & 239.53             \\
                 & Split Ways~\cite{khan2023split} & 4          & 85.31         & 50318             & 37840              \\
                 & Plaintext   & 4          & 88.06         & 8.56              & 0.033              \\
\textbf{PTB-XL}  & HESplit  & 4          & 58.71         & 100060.60         & 2624.85            \\
                 & Split Ways~\cite{khan2023split} & 4          & 65.42         & 72534             & 115640             \\
                 & Plaintext   & 4          & 67.68         & 15.55             & 0.317              
\end{tblr}
}
\end{table}

\noindent \textbf{Open Science and Reproducible Research:} 
To support open science and reproducible research, and provide researchers with the opportunity to use, test, and hopefully extend our work, our source code has been made available online\footnote{\href{https://github.com/khoaguin/HESplitNet}{https://github.com/khoaguin/HESplitNet}}.

\section{Conclusion}
\label{sec:conclusion}
Split learning is a collaborative learning technique that has been proposed as a privacy-preserving technique. Split learning-based solutions have been used in both open-source and commercial applications. However, recent studies have indicated that this approach is not infallible. Hence, it is important to consider other privacy-preserving techniques in order to ensure the security of the data. In this research, we demonstrate how the privacy leakage in split learning can be reduced using a different privacy-preserving technique -- homomorphic encryption\footnote{There is a plethora of other PPML techniques utilizing HE but to the best of our knowledge, until now there is only one work that combines SL with encrypted data.}. 
In homomorphic encryption, the computation is performed on encrypted data (which is computationally expensive). 
While this work features a single fully connected layer on the server side due to the constraints of encrypted training with HE, we acknowledge the potential for further expansion in future work. In upcoming research, we plan to explore advanced privacy enhancing technologies to incorporate more complex layers on the server side, balancing efficiency and model complexity. Furthermore, this work only considers a one-client setting, leaving the multi-client setting as a future work.

\newpage

\bibliographystyle{splncs04}
\bibliography{mybibliography}

\appendix

\end{document}